\documentclass[apj]{emulateapj}
\usepackage{amsmath}
\usepackage{blkarray}
\usepackage{graphicx}
\usepackage[usenames,dvipsnames]{color}

\graphicspath{{img/}}
\newcommand{\MWO}{{\rm MWO}}
\newcommand{\SSS}{{\rm SSS}}
\newcommand{\KKL}{{\rm KKL}}
\newcommand{\RpHK}{R'_{\rm HK}}

\newcommand{\cyc}{{\rm cyc}}
\newcommand{\cmin}[1]{{\rm #1, min}}
\newcommand{\cmax}[1]{{\rm #1, max}}
\newcommand{\ncmin}{{\rm min}}
\newcommand{\ncmax}{{\rm max}}

\newcommand{\anglemean}[1]{\ensuremath{\langle #1 \rangle}}
\newcommand{\fluxunits}{erg cm$^{-2}$ s$^{-1}$}
\newcommand{\II}{{\sc \romannumeral 2}}

\makeatletter
\renewcommand*\env@matrix[1][*\c@MaxMatrixCols c]{%
  \hskip -\arraycolsep
  \let\@ifnextchar\new@ifnextchar
  \array{#1}}
\makeatother

\newcommand{\revone}[1]{{#1}} 

\defcitealias{IAU:2015:B3}{IAU General Assembly 2015 Resolution B3}

\begin{document}

\title{The Mount Wilson Observatory $S$-index of the Sun}

\author{Ricky Egeland\altaffilmark{1,2}, Willie Soon\altaffilmark{3}, Sallie Baliunas\altaffilmark{4}, Jeffrey C. Hall\altaffilmark{5}, Alexei A. Pevtsov\altaffilmark{6,7}, and Luca Bertello\altaffilmark{8}}

\altaffiltext{1}{High Altitude Observatory, National Center for Atmospheric Research‡, PO Box 3000, Boulder, CO 80307-3000, USA; \email{\mbox{egeland@ucar.edu}}}
\altaffiltext{2}{Department of Physics, Montana State University, Bozeman, MT 59717-3840, USA}
\altaffiltext{3}{Harvard-Smithsonian Center for Astrophysics, Cambridge, MA 02138, USA}
\altaffiltext{4}{No affiliation}
\altaffiltext{5}{Lowell Observatory, 1400 West Mars Hill Road, Flagstaff, AZ 86001, USA}
\altaffiltext{6}{National Solar Observatory, Sunspot, NM 88349, USA}
\altaffiltext{7}{ReSoLVE Centre of Excellence, Space Climate research unit, 90014 University of Oulu, Finland}
\altaffiltext{8}{National Solar Observatory, Boulder, CO 80303, USA}

\begin{abstract}
The most commonly used \revone{index} of stellar magnetic activity is the
instrumental flux scale of \revone{singly-ionized} calcium H \& K line core emission,
$S$, developed by the Mount Wilson Observatory (MWO) HK Project, or
the derivative index $\RpHK$.  Accurately placing the Sun on the $S$
scale is important for comparing solar activity to that of the
Sun-like stars.  We present previously unpublished measurements of the
reflected sunlight from the Moon using the second-generation MWO HK
photometer during solar cycle 23 and determine cycle minimum
\revone{$S_\cmin{23} = 0.1634 \pm 0.0008$}, amplitude $\Delta S_{23} = 0.0143 \pm
0.0012$, and mean $\anglemean{S_{23}} = 0.1701 \pm 0.0005$.  By
establishing a proxy relationship with the closely related National
Solar Observatory Sacramento Peak calcium K emission index, itself
well-correlated with the Kodaikanal Observatory plage index, we
extend the MWO $S$ time series to cover cycles 15--24 and find on
average \revone{$\anglemean{S_\ncmin} = 0.1621 \pm 0.0008$,
  $\anglemean{\Delta S_\cyc} = 0.0145 \pm 0.0012$}, $\anglemean{S_\cyc} = 0.1694
\pm 0.0005$.  Our measurements represent an improvement over previous
estimates which relied on stellar measurements or solar proxies with
non-overlapping time series.  We find good agreement from these
results with measurements by the Solar-Stellar Spectrograph at Lowell
Observatory, an independently calibrated instrument, which gives us
additional confidence that we have accurately placed the Sun on the
$S$-index flux scale.
\end{abstract}


\section{Introduction}

Solar magnetic activity rises and falls in a roughly 11-year cycle
that has been diligently measured with sunspot counts for over 400
years.  Mechanical (i.e. magnetic) heating in chromospheric plage
regions on the Sun leads to emission in the cores of the absorption
lines Ca \II{} H \& K \citep{Linsky:1970,Athay:1970}.  The correlation
between HK emission and magnetic flux on the Sun
\citep{Skumanich:1975,Harvey:1999,Pevtsov:2016}, allows the study of
magnetic variability on other stars, placing the Sun and its solar
cycle in context.  Olin Wilson's HK Project at the Mount Wilson
Observatory (MWO) regularly observed the Ca \II{} H \& K emission for a
sample of over 100 bright dwarf stars from early F to early M type
beginning in 1966, for the first time characterizing long-term
magnetic variability of stars other than the Sun \citep{Wilson:1978}.
A large body of work is derived from these observations which forms
the basis of our understanding on the relationship between magnetic
activity and variability on fundamental stellar properties \citep{Hall:2008}.  For
example, \cite{Noyes:1984} established that activity decreases with
rotation rate and is influenced by stellar convection, with deeper
convective zones resulting in higher activity.  Sensitive differential
photometry revealed the relationship between variation at visible
wavelengths and magnetic activity, with the amplitude of photometric
variability increasing with activity, as well as the separation of
stars into high-activity \emph{spot-dominated} and low-activity
\emph{faculae-dominated} photometric variability classes according to
the sense of the correlation between visible photometry and activity
\citep{Lockwood:1997, Radick:1998, Lockwood:2007}.  Using 25 years of
MWO data, \cite{Baliunas:1995} revealed the patterns of long-term
magnetic variability in the Olin Wilson sample, finding cycling, flat,
and irregularly variable stars.  These results have been used to
constrain and inform theoretical studies of solar and stellar dynamos,
giving crucial information on the sensitivity of the dynamo to
fundamental properties such as mass and rotation \citep{Soon:1993,
  Baliunas:1996b, Saar:1999, Bohm-Vitense:2007, Metcalfe:2016}.

Each of the above results places the Sun -- the only star in which
\revone{spatially} resolved observations in a variety of bandpasses are
available -- in a stellar context.  However, the usefulness of the
solar-stellar comparison is only as good as the accuracy of the Sun's
placement on the stellar activity scale.  The magnetic activity proxy
established by the MWO HK project is the $S$-index:

\begin{equation}\label{eq:S}
  S = \alpha \frac{N_H + N_K}{N_R + N_V}
\end{equation}

\noindent where $N_H$ and $N_K$ are the counts in 1.09 \AA{}
triangular bands centered on Ca \II{} H \& K in the HKP-2
spectrophotometer, and $N_R$ and $N_V$ are 20 \AA{} reference
bandpasses in the nearby continuum region, and $\alpha$ is a
calibration constant \citep{Vaughan:1978}.

The HKP-2 instrument is distinct from the coud\'{e} scanner used by
Olin Wilson at the 100-inch telescope at MWO, later designated HKP-1
in \cite{Vaughan:1978}.  HKP-1 was a two-channel photometer, with one
1 \AA{} channel centered on either the H- or K-line and the other
channel measuring two 25 \AA{} bands separated by about 250 \AA{} from
the HK region \citep{Wilson:1968}.  HKP-1 measurements were therefore:

\begin{align}\label{eq:F}
\begin{split}
  F_H & = \frac{N_H}{N_\mathcal{R} + N_\mathcal{V}} \\
  F_K & = \frac{N_K}{N_\mathcal{R} + N_\mathcal{V}} \\
  F   & = \tfrac{1}{2} ( F_H + F_K )
\end{split}
\end{align}

where we use $\mathcal{R}$ and $\mathcal{V}$ to distinguish the
difference between the reference channels in the two instruments, with
HKP-1 $\mathcal{R}$ and $\mathcal{V}$ being 5 \AA{} wider than HKP-2
$R$ and $V$.  The $\alpha$ parameter of equation \eqref{eq:S} was
determined nightly with the standard lamp and standard stars such that
\emph{on average} $S = F$ \citep{Vaughan:1978,Duncan:1991}.  However,
differences are expected given that the two instruments are not
identical, and \cite{Vaughan:1978} derived the following relation with
coincident measurements on 13 nights in 1977:

\begin{equation}\label{eq:F2S}
  F = 0.033 + 0.9978 \, S - 0.2019 \, S^2
\end{equation}

It is important to stress that $S$ is an \emph{instrumental} flux scale
of the HKP-2 spectrophotometer that cannot be independently measured
without cross-calibration using overlapping Mount Wilson targets.  The
consequence of this is that there are only two methods of placing the
solar activity cycle on the $S$-index scale: directly measuring solar
light with the HKP-2 instrument, or calibrating another measurement to
the $S$-index scale using some proxy.  Previously, only the latter
method has been possible.  In this work we analyze hitherto
unpublished observations of the Moon with the HKP-2 instrument and
determine the placement of the Sun on the $S$-index scale.  We review
past calibrations of solar $S$ in section \ref{sec:prevS}.  In
section \ref{sec:obs} we describe the observations used in the
determination of solar $S$, and the analysis procedure in section
\ref{sec:analysis}.  In section \ref{sec:ks_linearity} we empirically
explore the assertion that $S$ is linear with Ca K-line emission.  We
conclude in section \ref{sec:conclusion} with a discussion on the
implications of our results, and future directions for establishing
the solar-stellar connection.

\section{Previous Solar $S$ Proxies}
\label{sec:prevS}

Sun-as-a-star Ca \II{} H \& K measurements have been made at Kitt Peak
National Observatory (NSO/KP) on four consecutive days each month from
1974--2013 \citep{White:1978}, and at Sacramento Peak (NSO/SP) daily, albeit with
with gaps, from 1976--2016. \citep{Keil:1984,Keil:1998}.  Results from
these observations for three solar cycles are given in
\cite{Livingston:2007}.  From the NSO/KP and NSO/SP spectrographs, the
K emission index (hereafter $K$) is computed as the integrated flux in
a 1 \AA{} band centered on the Ca \II{} K-line normalized by a band in
the line wing.  The principal difference of $K$ with respect to $S$ is
(1) $K$ includes flux only in the K line, while $S$ measures flux in
both H and K, (2) the reference bandpass is in the line wing for $K$,
as opposed to two 20 \AA{} bands in the nearby pseudo-continuum region
for $S$.  These differences could be cause for concern relating $S$ to
$K$, however (1) H and K are are a doublet of singly ionized calcium and
thus are formed by the same population of excited ions, therefore the
ratio $K/H$ cannot vary except by changes in the optical depth of the
emitting plasma \citep{Linsky:1970}, which is unlikely to vary by a
large amount (2) the far wing of the K line does vary somewhat over
the solar cycle, however only to a level of 1\% \citep{White:1981}.
Therefore we can expect \emph{a priori} that there should be a simple
linear relationship between the $K$ and and $S$ indices.  We
investigate this assumption in detail in Section
\ref{sec:ks_linearity}.

\begin{deluxetable*}{llcccc}
  \tabletypesize{\footnotesize}
  \tablecolumns{6} 
  \tablewidth{0pt}
  \tablecaption{ $S(K)$ Transformations  \label{tab:lit_k2s}}
  \tablehead{\colhead{Reference} & \colhead{$S(K)$} & \colhead{$S_\cmin{23}$} & \colhead{$S_\cmax{23}$} & \colhead{$\Delta S_{23}$} & \colhead{$\anglemean{S_{23}}$} }
  \startdata
  \cite{Duncan:1991}          & $(1.58 \pm 0.33) \, K + (0.040 \pm 0.002)$ $^a$ & 0.179 & 0.194 & 0.0151 & 0.187 \\[0.5em]
  \cite{White:1992}, original & $1.69 \, K + 0.016$ $^b$                        & 0.165 & 0.181 & 0.0162 & 0.173 \\[0.5em]
  \cite{White:1992}, mean     & $(1.64 \pm 0.07) \, K + (0.028 \pm 0.007)$ $^b$ & 0.172 & 0.188 & 0.0156 & 0.180 \\[0.5em]
  \cite{Baliunas:1995}        & $2.63 \, K - 0.066$ $^c$                        & 0.166 & 0.191 & 0.0251 & 0.178 \\[0.5em]
  \cite{Radick:1998}          & $(1.475 \pm 0.070) \, K + (0.041 \pm 0.013)$    & 0.171 & 0.185 & 0.0141 & 0.178 \\[0.5em]
  \cite{Hall:2004}            & $1.359 \, K + 0.0423$ $^d$                      & 0.162 & 0.175 & 0.0130 & 0.168 \\[0.2em]
  \hline \\[-0.5em]
  This Work                   & \revone{$(1.50 \pm 0.13) \, K + (0.031 \pm 0.013)$}      & 0.163 & 0.178 & 0.0143 & 0.170
  \enddata
  \tablecomments{$^a$: Duncan's $r_{\rm HK}$ replaced by $r_{\rm K}
    \equiv K$ using $r_{\rm HK}/r_{\rm K} = 0.089/0.087$ from their
    paper. Uncertainties are from a formal linear regression done in
    \cite{White:1992}, who noted the slope uncertainty is
    unrealistically large.  $^b$: Original calibration used the NSO/KP
    measurements; this version is transformed to use the NSO/SP
    measurements using Equation \ref{eq:sac2kitt} $^c$: Calculated
    using the published intervals and $\anglemean{S}$ in
    \cite{Donahue:1995}, along with the NSO/SP $K$ data. $^d$:
    Calculated from the published yearly mean values of $S$ at cycle
    23 minimum and maximum.}
\end{deluxetable*}

Several authors have developed transformations from the $K$ to MWO
$S$-index, and the results are summarized in Table \ref{tab:lit_k2s}.  The
earliest attempt was from \cite{Duncan:1991}, who used spectrograms of
16 MWO stars taken between 1964 and 1966 at the coud\'{e} focus of the
Lick 120 inch telescope to estimate a stellar $K$, and thereby establish a
relationship with an average of $S$ for those stars from MWO.  The Sun
was also a data point used in determining the relationship, with its
$K$ determined by NSO/KP and $S$ from \cite{Wilson:1978}'s
observations of the Moon at cycle 20 maximum and cycle 20-21 minimum.
\cite{Radick:1998} revisited the \cite{Duncan:1991} calibration using
longer time averages of $S$ for the stars and $K$ for the Sun with
updated observations, and adjusting the zero-point of the regression
to force the Sun's residual to zero.  These approaches neglect the
potential difference in scaling among the NSO/SP or NSO/KP $K$-indices
and the $K$-index derived from the Lick Spectrograph, but such a cross
correlation is not feasible in any case, due to the lack of common
targets for the different spectrographs.  Furthermore, this method
only uses a single measurement of the $K$-index for the stellar
sample, while the $S$-index is a decades-long average from Mount
Wilson data.  This poor sampling of $K$ will result in large scatter
due to rotational and cycle-scale activity for the active stars in the
sample, increasing the uncertainty in the determination of the $S(K)$
scaling relation.

\cite{White:1992} took another approach, leveraging Olin Wilson's
observations of the Moon with the original HKP-1 instrument
during cycle 20 \citep[][Table 3]{Wilson:1978}. However, because the
HKP-1 Moon observations did not overlap with the NSO $K$-index programs,
those time series had to be projected back in time using an intermediary solar
activity proxy, the 10.7 cm radio flux measurements (hereafter abbreviated
$F_{10.7}$).  \cite{White:1992} was discouraged that the result
from this method was discrepant with the \cite{Duncan:1991}
calibration, and citing the validity of both approaches, they chose to
average the two results.

The \cite{White:1992} $S(K)$ relationship was based on the NSO/KP
data, which is on a slightly different flux scale than the NSO/SP data
we use in this work.  \cite{White:1998} determined a linear
relationship between the two instruments to be $K_{\rm KP} = 1.1
\, K_{\rm SP} - 0.01$.  Using a cycle shape model fit (see section
\ref{sec:23fit}), we determined the cycle minima preceding cycles
22, 23, and 24, and the maxima of cycles 21, 22, and 23 in both
data sets.  Then, using a ordinary least squares regression on these
data, we obtained:

\begin{equation}
  K_{\rm KP} = 1.143 \, K_{\rm SP} - 0.0148
  \label{eq:sac2kitt}
\end{equation}

\noindent which is in agreement with the \cite{White:1998} relationship to the
precision provided.  We substitute \eqref{eq:sac2kitt} in the
\cite{White:1992} original and mean transformations and the results
are shown in Table \ref{tab:lit_k2s}.  Note that in the mean with
\cite{Duncan:1991} transformation we do \emph{not} use equation
\eqref{eq:sac2kitt} in the latter, as it was determined from stellar
observations using the Lick spectrograph, and therefore not specific
to NSO/KP data.

\cite{Baliunas:1995} observed that the \cite{White:1992} result failed
to cover the cycle 20-21 minima values of \cite[][Table
  3]{Wilson:1978}. Furthermore, they noted that their calibration
resulted in maxima for cycles 21 and 22 that were approximately equal
to the amplitude of cycle 20 measured by Wilson with HKP-1, while
other activity proxies (the sunspot record and  $F_{10.7}$) show cycle 20 to be significantly
weaker than cycles 21 and 22.  With these problems in mind,
\cite{Baliunas:1995} derived a new transformation $S(K)$ which
smoothed the cycle 20 to 21 minima transition from the Wilson
measurements to the $S(K)$ proxy and preserved the relative amplitudes
found in $F_{10.7}$ and sunspot records.  This transformation was not
published, however it was used again in \cite{Donahue:1995} who
published mean $S$ values from this transformation for several
intervals.  Using these intervals and mean values along with the
NSO/SP record we computed the \cite{Baliunas:1995} $S(K)$
relationship, which is shown in Table \ref{tab:lit_k2s}.

The Solar-Stellar Spectrograph (SSS) at Lowell Observatory
synoptically observes the Ca H \& K lines for $\sim$100 FGK stars, as
well as the Sun \citep{Hall:1995,Hall:2007b}.  The spectra are placed
on an absolute flux scale, and the MWO $S$-index is determined using
an empirically calibrated relationship \citep{Hall:1995}.  The
calibration is shown to be consistent with actual MWO observations to
a level of 7\% rms for low-activity stars \citep[see
  Figure 4][]{Hall:2007b}.  In \cite{Hall:2004}, mean values of $S$ for
cycle 23 minimum and maximum were published, which we used with the
NSO/SP data to derive an $S(K)$ transformation, shown in Table
\ref{tab:lit_k2s}. \cite{Hall:2004} remarked that their cycle 23 $S$
amplitude was ``noticeably less'' than the \cite{Baliunas:1995}
amplitude for the stronger cycle 22 (as evidenced from other proxies,
such as the sunspot record).  \revone{Subsequent re-reductions of
  the SSS solar time series in \cite{Hall:2007b} and \cite{Hall:2009}
  reported mean values of 0.170 and 0.171 for cycle 23 observations,
  $\sim 5$\% lower than the mean value of 0.179 reported in
  \cite{Baliunas:1995} for cycles 20--22.}

We used the cycle shape model fit (see section \ref{sec:analysis}) to
determine the values of the cycle 22-23 minima and cycle 23 maxima in
the NSO/SP $K$ timeseries.  We also computed the mean value
$\anglemean{K}$ of cycle 23 from this data set.  The conversion of
these $K$ values to $S$ using the previously published relationships
is shown in Table \ref{tab:lit_k2s}.  Each relationship arrives at
different conclusions about the placement of solar minimum, maximum,
and mean value for this cycle.  The relative range (max - min)/mean of
minima positions is $\approx$ 10\%, amplitudes $\approx$ 81\%, and
cycle means $\approx$ 11\%.  These discrepancies are significantly
larger than the uncertainty of the determination of these values from
$K$, which we estimate has an daily measurement uncertainty of $\sim$
1\% (see Section \ref{sec:analysis}) and a cycle amplitude of 6\%
above minimum.  The largest discrepancy, as \cite{Hall:2004} noted, is
in the cycle amplitude, with the \cite{Baliunas:1995} $S(K)$ proxy
estimate being more than double their measurement.  Considering the
variety and magnitude of these discrepancies, we conclude that
the solar $S$-index has so far not been well understood.

\section{Observations}
\label{sec:obs}

\subsection{Mount Wilson Observatory HKP-1 and HKP-2}
The Mount Wilson HK Program observed the Moon with both the HKP-1 and
HKP-2 instruments.  After removing 11 obvious outliers there are 162
HKP-1 observations taken from 2 Sep 1966 to 4 Jun 1977 with the Mount
Wilson 100-inch reflector, covering the maximum of cycle 20 and the
cycle 20-21 minimum.  \cite{Wilson:1968} and \cite{Duncan:1991}
published mean values from these data, with the latter shifted upward
by about 0.003 in $S$.  Our HKP-1 data is under the same calibration
as in \cite{Duncan:1991} and \cite{Baliunas:1995}.

As mentioned in \cite{Baliunas:1995}, observations of the Moon resumed
in 1993 with the HKP-2 instrument.  After removing 10 obvious outliers
there are 75 HKP-2 observations taken from 27 Mar 1994 to 23 Nov 2002
with the Mount Wilson 60-inch reflector, covering the end of cycle 22
and the cycle 23 minimum, extending just past the cycle 23 maximum.
The end of observations coincides with the unfortunate termination of
the HK Project in 2003.  These observations were calibrated in the
same way as the stellar HKP-2 observations as described in
\cite{Baliunas:1995}, using the standard lamp and measurements from
the standard stars.  Long-term precision of the HKP-2 instrument was
shown to be 1.2\% using 25 years of observations in a sample of 13
stable standard stars.

The 75 HKP-2 lunar observations are the only observations of solar
light with the HKP-2 instrument, and thus are the only means of
directly placing the Sun on the instrumental $S$-index scale of
equation \eqref{eq:S}.  We assume that these observations measure $S$
for the Sun to within the 1.2\% precision determined for the HKP-2
instrument.  Most HKP-1 and HKP-2 observations were taken within 5
days of full Moon, and all observations were taken within an hour of
local midnight.  We do not expect a significant alteration of the
nearby spectral bands constituting $S$ by reflection from the Moon,
which is to first order a gray diffuse reflector.

\subsection{NSO Sacramento Peak K-line Program}

We seek to extend our time series of solar variability beyond cycle 23
by establishing a proxy to the NSO Sacramento Peak (NSO/SP)
observations\footnote{\url{ftp://ftp.nso.edu/idl/cak.parameters}}
taken from 1976--2016, covering cycles 21 to 24.  The NSO/SP K-line
apparatus is described in \cite{Keil:1984}.  Briefly, it consists of a
$R \sim 150,000$ Littrow spectrograph installed at the John W. Evans
Solar Facility (ESF) at Sacramento Peak Observatory, fed by a
cylindrical objective lens which blurs the Sun into a 50 $\mu$m by 10
mm line image at the spectrograph slit.  The spectral intensity scale
is set by a integrating a 0.53 \AA{} band centered at 3934.869 \AA{}
in the K-line wing and setting it to the fixed value of 0.162.  The
$K$ emission index is then defined as the integrated flux of a 1 \AA{}
band centered at the K-line core (3935.662 \AA{}) \citep{White:1978}.
We estimated the measurement uncertainty to be 1.0\% by calculating
the standard deviation during the period 2008.30--2009.95, which is
exceptionally flat in the sunspot record and 10.7 cm radio flux time
series.

\subsection{Kodaikanal Observatory Ca K Spectroheliograms}

We extend the $S$-index record back to cycle 20 using the composite
$K$ time series of \cite{Bertello:2016}, which is available online
\citep{Data:Kcomposite}.  This composite calibrates the NSO/SP data to
K-line observations by the successor program at NSO, the Synoptic Optical
Long-term Investigations of the Sun (SOLIS) Integrated Sunlight
Spectrometer (ISS) \citep{Bertello:2011}.  The
calibration used \revone{in \cite{Bertello:2016}} is:

\begin{equation}\label{eq:sac2iss}
  K_{\rm ISS} = 0.8781 \, K_{\rm SP} + 0.0062
\end{equation}

\revone{\cite{Bertello:2016} calibrated the synoptic Ca \II{} K plage
  index from spectroheliograms from the Kodaikanal (KKL) Observatory
  in India to the ISS flux scale using the overlapping portion NSO/SP
  data, resulting in a time series of Ca \II{} K emission from 1907 to
  the present.  We transform this composite timeseries from the ISS
  flux scale to the NSO/SP flux scale by applying the inverse of
  equation \eqref{eq:sac2iss}:

\begin{equation}\label{eq:iss2sp}
  K_{\rm KKL(SP)} = 1.1388 \, K_{\rm  KKL(ISS)} - 0.0071
\end{equation}
}

We prefer this homogeneous chromospheric
K-line proxy (hereafter denoted simply $K_\KKL$) over proxies based on
photospheric phenomena such as $F_{10.7}$ or the sunspot number.  In
particular, \cite{Pevtsov:2014} found that the correlation between $K$
and $F_{10.7}$ is non-linear and varies with the phase of the solar
cycle, with strong correlation during the rising and declining phases,
and poor correlation at maximum and minimum, precisely the sections of
the solar cycle of most interest in this work.

\subsection{Lowell Observatory Solar Stellar Spectrograph: Updated
  Data Reduction}
\label{sec:obs:sss}

We compare our results to \revone{a new reduction of} observations
from the Lowell Observatory Solar-Stellar Spectrograph (SSS), which is
running a long-term stellar activity survey complementary to the MWO
HK Project.  The SSS observes solar and stellar light with the same
spectrograph, with the solar telescope consisting of an exposed
optical fiber that observes the Sun as an unresolved source
\citep{Hall:1995,Hall:2007b}.  The basic measurement of SSS is the
integrated flux in 1 \AA{} bandpasses centered on the Ca \II{} H \& K
cores from continuum-normalized spectra, $\phi_{\rm HK}$, which can
then be transformed to the $S$-index using a combination of empirical
relationships derived from stellar observations:

\begin{equation}\label{eq:S_SSS}
  S_\SSS = \frac{10^{14} \mathcal{F}_{\rm c, \lambda 3950}}{K_\mathcal{F} C_{\rm cf} T_{\rm eff}^4} \phi_{\rm HK}
\end{equation}

\noindent where $\mathcal{F}_{\rm c, \lambda 3950}$ is the continuum
flux scale for the Ca \II{} H \& K wavelength region, which converts
$\phi_{\rm HK}$ to physical flux (\fluxunits{}).  $\mathcal{F}_{\rm c,
  \lambda 3950}$ is a function of Str\"{o}mgren $(b - y)$ and is taken
from \cite{Hall:1996}.  $K_\mathcal{F}$ (simply $K$ in other works) is
the conversion factor between the MWO HKP-2 H \& K flux (numerator of
equation \eqref{eq:S}) to physical flux \citep{Rutten:1984}.  $C_{\rm
  cf}$ a factor that removes the color term from $S$, and is a
function of Johnson $(B-V)$ \citep{Rutten:1984}.  Finally, $T_{\rm
  eff}$ is the effective temperature.  See \cite{Hall:2007b} and
\cite{Hall:1995} for a details on the extensive work leading to this
formulation.  What is important to realize about this method of
obtaining $S$ is that it requires three measurements of solar
properties, $(b - y)_\Sun$, $(B-V)_\Sun$, and $T_{\rm eff,\Sun}$,
along with the determination of one constant, $K_\mathcal{F}$.  The
solar properties are taken from best estimates in the literature,
which vary widely depending on the source used, and can dramatically
affect the resulting $S_\SSS$ for the Sun.  \cite{Hall:2007b} used $(b
- y)_\Sun = 0.409$, $(B-V)_\Sun = 0.642$, and $T_{\rm eff,\Sun} =
5780$ K.  The constant $K_\mathcal{F}$ was empirically determined to
be $0.97 \pm 0.11$ \fluxunits{} in \cite{Hall:2007b} as the value
which provides the best agreement between $S_\SSS$ and $S_\MWO$ from
\cite{Baliunas:1995} for an ensemble of stars and the Sun.  This
combination of parameters resulted in a mean $S_\SSS$ of 0.170 for the
Sun using observations covering cycle 23.  A slightly different calibration of SSS data in
\cite{Hall:2004} used a flux scale $\mathcal{F}_{\rm c, \lambda 3950}$
based on Johnson $(B-V)$, set to 0.65 for the Sun, and $T_{\rm
  eff,\Sun} = 5780$ K.  In Table \ref{tab:lit_k2s} we estimated that
this calibration resulted in a mean $S = 0.168$ for cycle 23.
\cite{Hall:2009}, which included a revised reduction procedure and one
year of data with the upgraded camera (see below), found
$\anglemean{S} = 0.171$.

As mentioned previously, the three solar properties $(b - y)_\Sun$,
$(B-V)_\Sun$, and $T_{\rm eff,\Sun}$ used in the SSS flux-to-$S$
conversion are not accurately known.  The fundamental problem is that
instruments designed to observe stars typically cannot observe the
Sun.  \cite{CayreldeStrobel:1996} studied this problem, and collected
$(B-V)_\Sun$ from the literature ranging from 0.62 to 0.69.
\cite{Melendez:2010} compiled literature values $(b - y)_\Sun$ ranging
from 0.394 to 0.425.  $T_{\rm eff, \Sun}$ is more accurately known,
which is fortunate given that it appears in equation \eqref{eq:S_SSS}
to the fourth power.  However, a 0.01 change in $(B-V)_\Sun$ or $(b -
y)_\Sun$ results in an approximately 2\% or 10\% change in $S_\SSS$,
respectively.  \revone{The sensitivity of $S_\SSS$ to these properties
makes it especially important to use the best known values.}

More recent photometric surveys of solar analogs have resulted in
improved determinations of the solar properties by way of
color-temperature relations.  We are therefore motivated to update
$S_\SSS$ for the Sun using these measurements: $(B - V)_\Sun = 0.653
\pm 0.003$ \citep{Ramirez:2012}, $(b - y)_\Sun = 0.4105 \pm 0.0015$
\citep{Melendez:2010}, and $T_{\rm eff, \Sun} = 5772.0 \pm 0.8$
\citepalias{IAU:2015:B3}.  The latter lower value for the effective
temperature follows from the recent lower estimate of total solar
irradiance in \cite{Kopp:2011}.  The constant $K_\mathcal{F}$ is kept
at 0.97 as determined in \cite{Hall:2007b}.  The SSS data analyzed
here now include data taken after upgrading the instrument CCD to an
Andor iDus in early 2008 \citep{Hall:2009}.  This new CCD has higher
sensitivity in the blue and reduced read noise.  The reduction
procedure remains the same as described in \cite{Hall:2007b} and on
the SSS web
site\footnote{\url{http://www2.lowell.edu/users/jch/sss/tech.php}},
albeit with updated software and two additional steps: (1) high S/N
spectra from the new camera are used as reference spectra for the old
camera data, which improves stability of the older data and avoids
discontinuity at the camera upgrade boundary (2) an additional scaling
correction is applied to the continuum-normalized spectra so that the
line wings are shifted to match the normalized intensity of the
\cite{Kurucz:1984} solar spectrum.  Despite these efforts, a
non-negligible discontinuity was apparent across the CCD upgrade
boundary.  We suspect this may be due to minute differences in CCD
pixel size, and the slightly different wavelength sampling at the
continuum normalization reference points.  The discontinuity is
corrected \emph{post facto} by multiplying the CCD-1 $S$ data by
0.9710, determined by the ratio of the medians for the last year of
CCD-1 data to the first year of CCD-2 data.  Finally, due to tape
degradation, raw CCD data from 1998-2000 were lost, preventing
reduction using the updated routines.  Continuum normalized spectra
from that period still exist, though introducing them into the updated
pipeline results in significant discontinuities in $S$ which had to be
corrected.  The correction consisted of applying an additional scaling
factor to the lost data region such that the region median falls on a
linear interpolation of the cycle across the region.  Keeping or
removing the ``corrected'' data in this region does not affect our
conclusions.  Finally, we estimated the measurement uncertainty to be
1.6\% for CCD-1, and 1.3\% for CCD-2, by computing the standard
deviation of observations from each device in the long minimum from
2007--2010.

\section{Analysis}
\label{sec:analysis}

Our goal in this work is to use the 75 HKP-2 observations to determine
the $S$-index for the Sun.  Specifically, we seek to measure the
minimum, maximum, and mean value of $S$ over several solar cycles.  We
proceed first by directly measuring these quantities using our cycle
23 HKP-2 data, and then establish a proxy with the NSO/SP measurements
to extend our measurements to other cycles.

\subsection{Cycle 23 Direct Measurements}
\label{sec:23direct}

\begin{figure*}[htb!]
  \centering
  \includegraphics[width=0.83\textwidth]{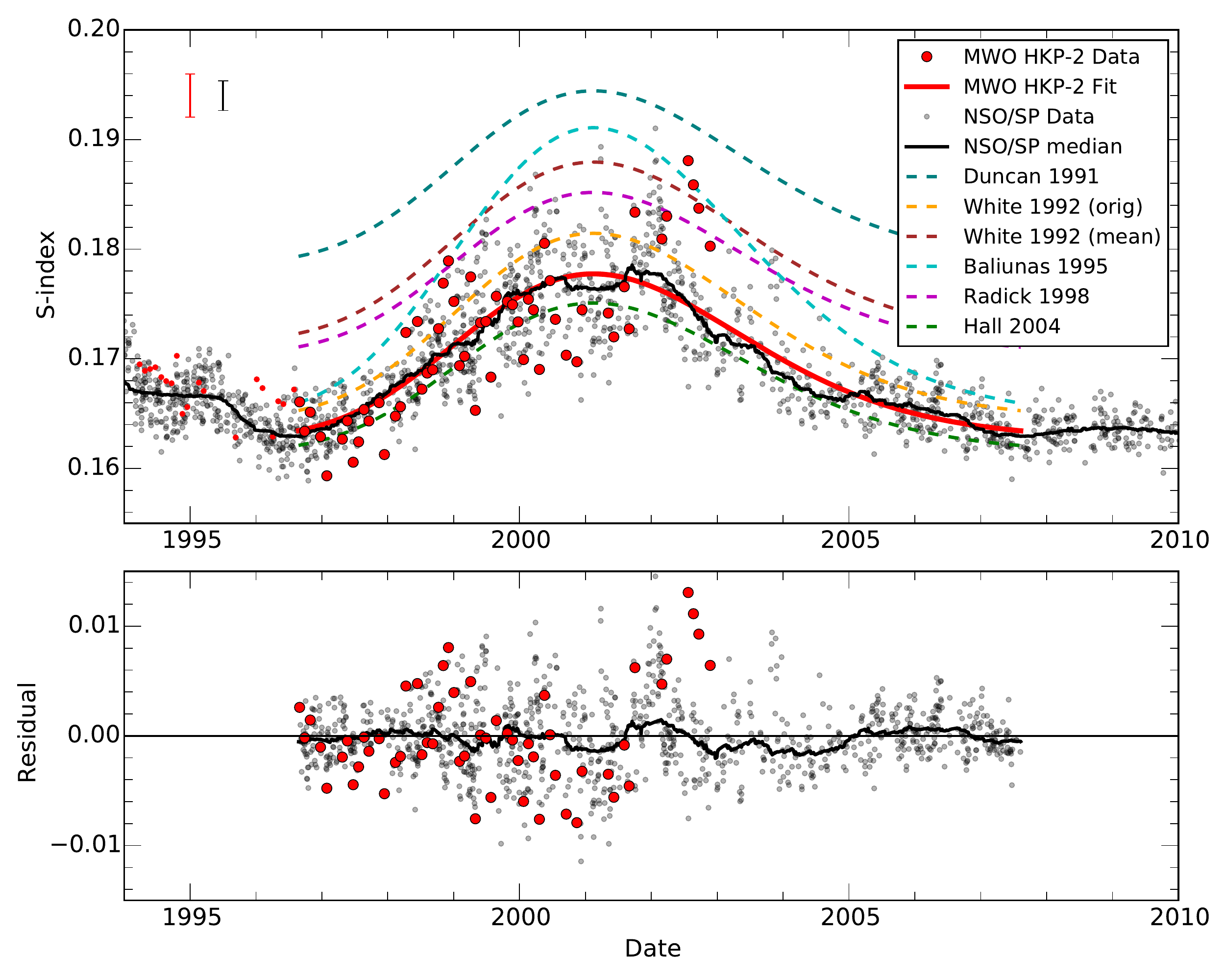}
  \caption{Cycle 23 data and cycle shape model fits.  The MWO HKP-2
    Moon observations are shown in red, with the larger points used in
    the cycle shape model fit, which is shown as a thick red curve.
    NSO/SP $K$-index data are transformed to the $S$ scale using
    equation \eqref{eq:k2s}.  A 1-yr wide median filter applied to the
    NSO/SP data is shown in the solid black line.  $S(K)$
    transformations of found in the literature (see Table
    \ref{tab:lit_k2s}) of a cycle shape model fit to the NSO/SP $K$
    data are shown as colored curves for comparison.  The bottom panel
    shows the residual difference of the cycle shape model curve and
    the data.  Error bars in the top left show the estimated
    measurement uncertainty for MWO HKP-2 (red) and NSO/SP (black).}
  \label{fig:cyc23}
\end{figure*}

The HKP-2 data can be used to measure the minimum, maximum and mean of
cycle 23 directly, but first the time of minimum and maximum must be
established by some other means.  We choose to use the NSO/SP $K$
record for this purpose.  Applying a 1-yr boxcar median filter to the
\revone{NSO/SP} time series, we find the absolute minimum between
cycles 22-23 at decimal year 1996.646, and absolute maximum at
2001.708.  \revone{In this interval defining cycle 23, there are 56
  HKP-2 measurements, with the remaining 19 points belonging to cycle
  22.  Next, using the HKP-2 time series,} we take the median of a 2-yr
wide window centered at \revone{the minima and maxima times} to find
$S_\cmin{23} = 0.1643$ ($N = 17$) and $S_\cmax{23} = 0.1755$ ($N =
12$), for a cycle 23 amplitude $\Delta S_{23} = 0.0112$.  The mean
value for the cycle $\anglemean{S_{23}} = 0.172$ (N = 56).  We choose
the median over the mean when measuring fractions of a cycle because
we do not expect Gaussian distributions in that case.  We choose a
2-yr wide window to be about as wide as we can reasonably go without
picking up another phase of the solar cycle.  Still, the number of
points in each window is low, which does not give much confidence that
the cycle can be \revone{precisely} measured in this way.  We shall
investigate the uncertainty of this method in the next section.  For
the cycle mean, besides the problem of low sampling, we would expect
this to be an overestimate since no data exists for the longer
declining phase of the cycle.

\revone{Nonetheless, even with} these simple estimates we find
discrepancies with the previous work shown in Table \ref{tab:lit_k2s}.
The measured minimum is appreciably lower than the \cite{Duncan:1991},
\cite{White:1992} mean, and \cite{Radick:1998} values, and the
amplitude is lower than all other estimates.  \revone{Our} cycle mean is also
lower than all but the \cite{Hall:2004} estimate, \revone{indicating
  that the previous work has overestimated the $S$-index of the Sun.}
In the next section we will attempt to improve our \revone{precision}
using a method in which more of the data are used.

\subsection{Cycle Shape Model Fit for Cycle 23}
\label{sec:23fit}

Due to the limited data we have from HKP-2 for cycle 23, the results
of the previous section are susceptible to appreciable uncertainties
from unsampled short-timescale variability due to rotation and active
region growth and decay.  By fitting a cycle model to our data, we can
reduce the uncertainty in our minimum and maximum point
determinations.  We use the skewed Gaussian cycle shape model of
\cite{Du:2011} for this purpose:

\begin{equation}\label{eq:cycshape}
  f(t) = A \exp\left( - \frac{(t - t_m)^2}{2 B^2 \left[ 1 + \alpha (t - t_m) \right]^2} \right) + f_\ncmin
\end{equation}

\noindent where $t$ is the time, $A$ is the cycle amplitude, $t_m$ is
approximately the time of maximum, $B$ is roughly the width of the
cycle rising phase, and $\alpha$ is an asymmetry parameter.
$f_\ncmin$, which was not present in the \citeauthor{Du:2011} model, is
an offset which sets the value of cycle minimum.  We also
tried the quasi-Planck function of \cite{Hathaway:1994} for this
purpose and obtained similar results, however we prefer the above
function due to the simplicity of interpreting its parameters and
developing heuristics to guide the fit.  \revone{We fit the model to
  the data using the Python {\tt scipy} library {\tt curve\_fit}
  routine with bounds.  This function uses a Trusted Region Reflective (TRR)
  method with the Levenberg-Marquardt (LM) algorithm applied to
  trusted-region subproblems \citep{Branch:1999,More:1978}.
 The fitting algorithm searches} for the optimum
parameters from a \revone{bounded} space within $\pm50$\% of heuristic values obtained
using a 1-yr median filter of the data, with the exception of
$f_\ncmin$ which has a lower bound set at the lowest data point in
order to prevent the fitting procedure from underestimating the
minimum.

The 56 \revone{HKP-2 data points contained in cycle 23} only cover the
rising phase and are not enough to constrain a least-squares fitting
procedure.  To circumvent this problem, we first fit the NSO/SP data
and assume that the parameters which determine the cycle shape $t_m$,
$B$, and $\alpha$ are the same in the two chromospheric time series.
The only way this assumption could fail is if the Ca \II{} H band, or
the 20 \AA{} continuum bands of $S$ (equation \ref{eq:S}) varied in
such a way as to distort the cycle shape with respect to $K$.
\cite{Soon:1993b} explored the long-term variability of the $C_{\rm
  RV}$ index based on the 20 \AA{} reference bands, finding it to be
generally quite small.  We therefore assume for the moment that the H,
R, and V bands are linear with $K$ or constant, and later confirm
these assumptions in Section \ref{sec:ks_linearity}.  Fitting equation
\eqref{eq:cycshape} to the NSO/SP $K$ for cycle 23, we find $t_m =
2001.122$, $B = 2.154$, and $\alpha = 0.0343$.  We then hold these
parameters fixed and fit equation \eqref{eq:cycshape} to the 56 HKP-2
observations for cycle 23, finding the remaining free parameters $A =
0.0150, f_\ncmin = 0.163$.

The cycle model fit and the HKP-2 data are shown as the red curve in
Figure \ref{fig:cyc23}.  The reduced $\chi^2$ of the fit is 6.45,
which we find acceptable given the model does not seek to explain all
the variation in the data (e.g. rotation, active region growth and decay), only
the mean cycle.  We find an RMS residual of the fit of $\sigma_{\rm
  res} = 0.0047$, which is a bit more than double the estimated individual
measurement uncertainty $\sigma = 0.0020$.

Using the cycle model fit, we find the minimum \revone{$S_\cmin{23} = 0.1634
\pm 0.0008$}, maximum $S_\cmax{23} = 0.1777 \pm 0.0010$, and \revone{amplitude
$\Delta S_{23} = 0.0143 \pm 0.0012$}.  The mean of the cycle model curve is
$\anglemean{S_{23}} = 0.1701 \pm 0.0005$.  The uncertainties are
determined from a Monte Carlo experiment in which we build a
distribution of cycle model fits from only 56 points of NSO/SP $K$
data during the rising phase of cycle 23, and comparing that to the
``true'' cycle model fit using all 1087 observations throughout the
cycle (see Appendix A for details).  The Monte Carlo experiment shows
that the cycle model fit method with only 56 randomly selected points
finds true minima and maxima with a $1\sigma$ standard deviation of
$\approx$0.5\%, while the method of direct means finds minima equally
well, but maxima with about double the uncertainty due to the
increased variability at that phase of the cycle.

Our results for cycle 23 \revone{are} shown in the final row of Table
\ref{tab:lit_k2s}, and are discrepant with most of the previous
literature values.  \cite{Hall:2004} comes closest to our result,
though their amplitude is 9\% lower.  The \cite{Radick:1998} amplitude
is only 0.002 $S$-units (1.4\%) lower than ours, but the mean is 4.7\%
higher due to the higher value for solar minimum.  The
\cite{Baliunas:1995} relation finds a minimum only 1.8\% higher than
ours, but the amplitude is substantially larger (75\%), leading to a
4.7\% higher estimate of the solar cycle mean.

The cycle shape model fit to the NSO/SP $K$ data is transformed using
the literature relations in Table \ref{tab:lit_k2s} and are shown as
colored curves in Figure \ref{fig:cyc23}.  Here we see that no
transformation matches the MWO HKP-2 observations in a satisfactory
way, though the \cite{Hall:2004} curve comes close.  In the next
section we will construct a new $S(K)$ transformation that exactly
matches our cycle shape model curve in Figure \ref{fig:cyc23}.

\subsection{$S(K)$ Proxy Using NSO/SP $K$ Emission Index}
\label{sec:k2s}

\revone{

We now seek a transformation between the NSO/SP $K$ emission index and
the Mount Wilson $S$-index.  We assume a linear relationship:

\begin{equation}\label{eq:k2s_model}
  S(K) = a + b \, K
\end{equation}

Now we write $K$ as a function of time using the cycle shape model
(equation \ref{eq:cycshape}), obtaining:

\begin{equation}
  S(t) = (a + b f_{\ncmin,K}) + b A_K \, E(t; t_m, B, \alpha)
\end{equation}

\noindent where $E(t; t_m, B, \alpha)$ is the exponential function
from \eqref{eq:cycshape}.  This is precisely the same form as
\eqref{eq:cycshape}, which is made more clear by defining $f_{\ncmin,S}
= (a + b f_{\ncmin,K})$ and $A_S = b A_K$.  These definitions may be
used as the solutions for $\{a, b\}$ when cycle shape model fits have
been done for both $K(t)$ and $S(t)$, as was done in the previous
section, giving

\begin{align}\label{eq:k2s_model_params}
\begin{split}
  a &= f_{\ncmin,S} - (A_S / A_K) \, f_{\ncmin,K} \\
  b &= A_S / A_K .
\end{split}
\end{align}

Note that the cycle shape parameters $\{t_m, B, \alpha\}$ are not
explicitly related to $\{a, b\}$.  Using the cycle 23 fit parameters
described in the previous section, we arrive at the following linear
transformation:

\begin{equation}\label{eq:k2s}
  S(K) = (1.50 \pm 0.13) K + (0.031 \pm 0.013)
\end{equation}

The uncertainty in the slope and intercept are calculated using
standard error propagation methods on equation
\eqref{eq:k2s_model_params}.

\begin{align}\label{eq:k2s_model_error}
\begin{split}
  \sigma_b^2 &= b^2 \, ( (\sigma_{A_S}/A_S)^2 + (\sigma_{A_K}/A_K)^2) \\
  \sigma_{f_K \cdot b}^2 &= (f_K \cdot b)^2 \cdot [ (\sigma_b / b)^2 \\
                         &+ (\sigma_{f_K} / f_K)^2 + 2 \, (\sigma_b / b) \, (\sigma_{f_K} / f_K) \,  \rho(f_K, b) ] \\
  \sigma_a^2 &= \sigma_{f_S}^2 + \sigma_{f_K \cdot b}^2 - 2 \, \sigma_{f_S} \, \sigma_{f_K \cdot b} \, \rho(f_S, f_K \cdot b)
\end{split}
\end{align}

\noindent where we have simplified the notation with $f_K \equiv
f_{\ncmin, K}$ and $f_S \equiv f_{\ncmin, S}$.  Correlation
coefficients $\rho(x, y)$ for quantities $x$ and $y$ are defined as
$\rho(x, y) = {\rm cov}(x, y)/(\sigma_x \cdot \sigma_y)$.  We
approximate $\rho(f_K, b) \sim \rho(f_K, A_K)$ and $\rho(f_S, f_{K
\cdot b}) \sim \rho(f_S, A_S)$, which likely overestimates the
correlation between these quantities.  The uncertainties of each of
the curve fit parameters are obtained from Monte Carlo experiments
described in Appendix A.  Correlation between $A_S$ and $A_K$ through
the cycle shape parameters $\{t_m, B, \alpha\}$ is
included in our estimate of $\sigma_{A_S}$ due to the setup of the
Monte Carlo experiment (see Appendix A).  The error budget is
dominated by $(\sigma_{A_S}/A_S)^2$, $(\sigma_b / b)^2$, and
$\sigma_{f_S}^2$, with all other terms accounting for less than 5\% of
the total in their respective equations.

The estimated uncertainty of the cycle shape model cross-calibration
method described above is significantly less than was achieved by
linear regression of coincident measurements.  The latter method had
formal uncertainties in the scale factor in excess of 100\%.  Linear
regression failed because we have few coincident measurements and the
individual measurement uncertainty of $\approx$ 1\% in both $K$ and
$S$ is roughly 10\% the amplitude of the variability over the cycle.

}

We transformed the NSO/SP $K$-index time series using equation
\eqref{eq:k2s} and plotted it with the MWO HKP-2 data and cycle shape
model fits in Figure \ref{fig:cyc23}.  A 1-yr boxcar median
filter of the data is also plotted as a black line, which is in good
agreement with the cycle shape model curve.

\subsection{Comparison with SSS Solar Data}
\label{sec:sss}

\begin{figure*}[htb!]
  \centering
  \includegraphics[width=\textwidth]{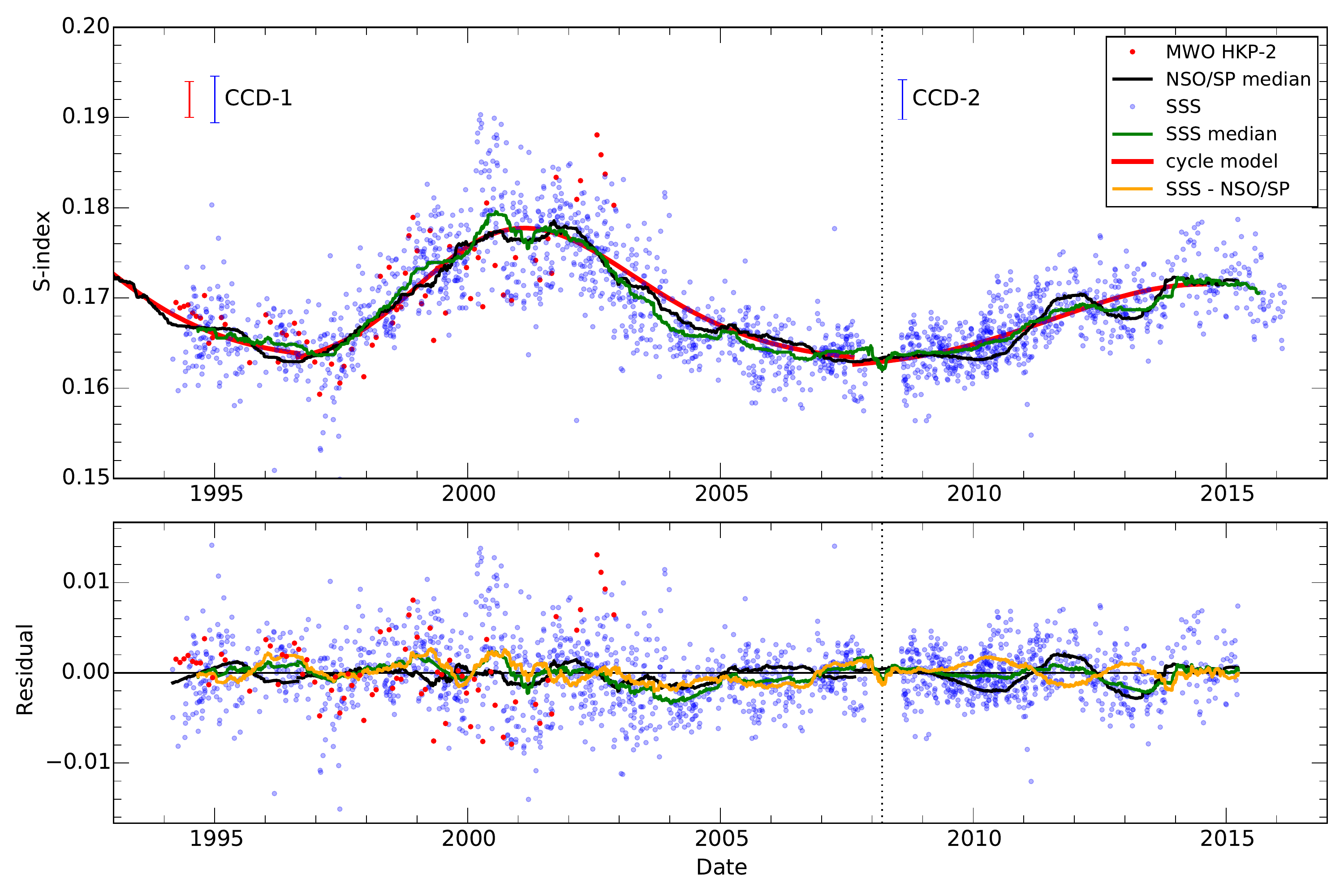}
  \caption{SSS solar observations (blue) compared to the MWO HKP-2
    measurements (red) and the NSO/SP $S(K)$
    proxy, shown here as a running 1-yr median (black).  Cycle
    shape model curves fit to the NSO/SP data are shown in red.
    The vertical dotted line denotes the upgrade of the SSS
    CCD.  The bottom panel shows the residual difference of the data
    with the cycle shape model, while the orange line is the
    difference between the SSS (green) and NSO/SP (black) running
    medians.  Error bars in the top panel show the estimated
    measurement uncertainty for SSS CCD-1 and CCD-2 observations
    (blue), and MWO HKP-2(red).}
  \label{fig:sss}
\end{figure*}

As an additional check of our transformation of the NSO/SP data to the
$S$-index scale, we compare our result to the independently calibrated
Solar-Stellar Spectrograph (SSS) observations \revone{described in
  section \ref{sec:obs:sss}.}

In Figure \ref{fig:sss} we overplot the Lowell Observatory SSS solar
$S$-index data on our NSO/SP $S(K)$ proxy for cycles 22, 23 and 24.  A 1-yr
median filter line is plotted for both time series, as well as
the cycle shape model curves \revone{fit to the NSO/SP $S(K)$ data using the TRR+LM
  algorithm described in section \ref{sec:23fit}}.  In
general, we find excellent agreement between all three curves.  We use
the cycle shape model curves as a reference to compare the NSO/SP
$S(K)$ proxy with SSS.  For cycle 23, which is covered by the SSS
CCD-1 data, the mean of the residual is -0.00031 $S$-units with a standard
deviation of 0.0040 $S$.  This may be compared to the NSO/SP data for the
same period used to constrain the model fit, with an essentially zero
mean and a standard deviation of 0.0032 $S$.  Similarly for cycle 24, SSS
CCD-2 data have a residual mean of -0.000097 $S$ and standard deviation of
0.0026, compared to the NSO/SP residual mean of -0.000047 $S$ and
standard deviation 0.0028 $S$.  In the case of cycle 24, SSS
observations have a lower residual with the cycle model than the
NSO/SP data used to define it!  The difference of the SSS and NSO/SP
running medians is shown as an orange line in Figure \ref{fig:sss},
and rarely exceeds $\pm 0.002$ $S$.

We now consider whether the remarkable agreement is confirmation of
the true solar $S$-index, or mere coincidence.  As discussed \revone{in
  section \ref{sec:obs:sss}}, the
computation of $S$ from SSS spectra is sensitive to the measurement of
solar color indices and the effective temperature.  We have
recalibrated the data with the best available measurements of these
quantities.  The resulting calibration results in a 3\% scaling
difference between CCD-1 and CCD-2 data, which we choose to remove by
rescaling CCD-1 data to the CCD-2 scale, which resulted in the
excellent agreement with NSO/SP $S(K)$.  There is good reason for this
choice, since CCD-2 is a higher quality detector.  However, \emph{if
  there were} a significant offset between the NSO/SP $S(K)$ and SSS,
we would be justified in applying a small scaling factor to reconcile
differences, citing uncertainties in the solar properties or the
conversion factor $K_\mathcal{F}$.  The fact that this was not
necessary might be considered coincidence.  However, the fact that
\emph{only} a scaling factor would be required, and not an absolute
offset, can \emph{not} be coincidence.  From our determination of the
$S(K)$ scaling relation \eqref{eq:k2s} using equation
\eqref{eq:k2s_model_params}, we see that using a different amplitude
\revone{$A_S$} would change the scale of the conversion, \revone{$b$,
  \emph{and} the offset, $a$}, which could not be confirmed by
these SSS data using a scaling factor alone.  Therefore, the agreement
between SSS and the NSO/SP $S(K)$ proxy can be taken as confirmation
of the latter.  The agreement between SSS and the MWO HKP-2 data
points (red points in Figure \ref{fig:sss}) is confirmation that
$S_\SSS$ is properly calibrated for the Sun.

\subsection{Calibrating HKP-1 Measurements}

\begin{figure*}[htb!]
  \centering
  \includegraphics[width=\textwidth]{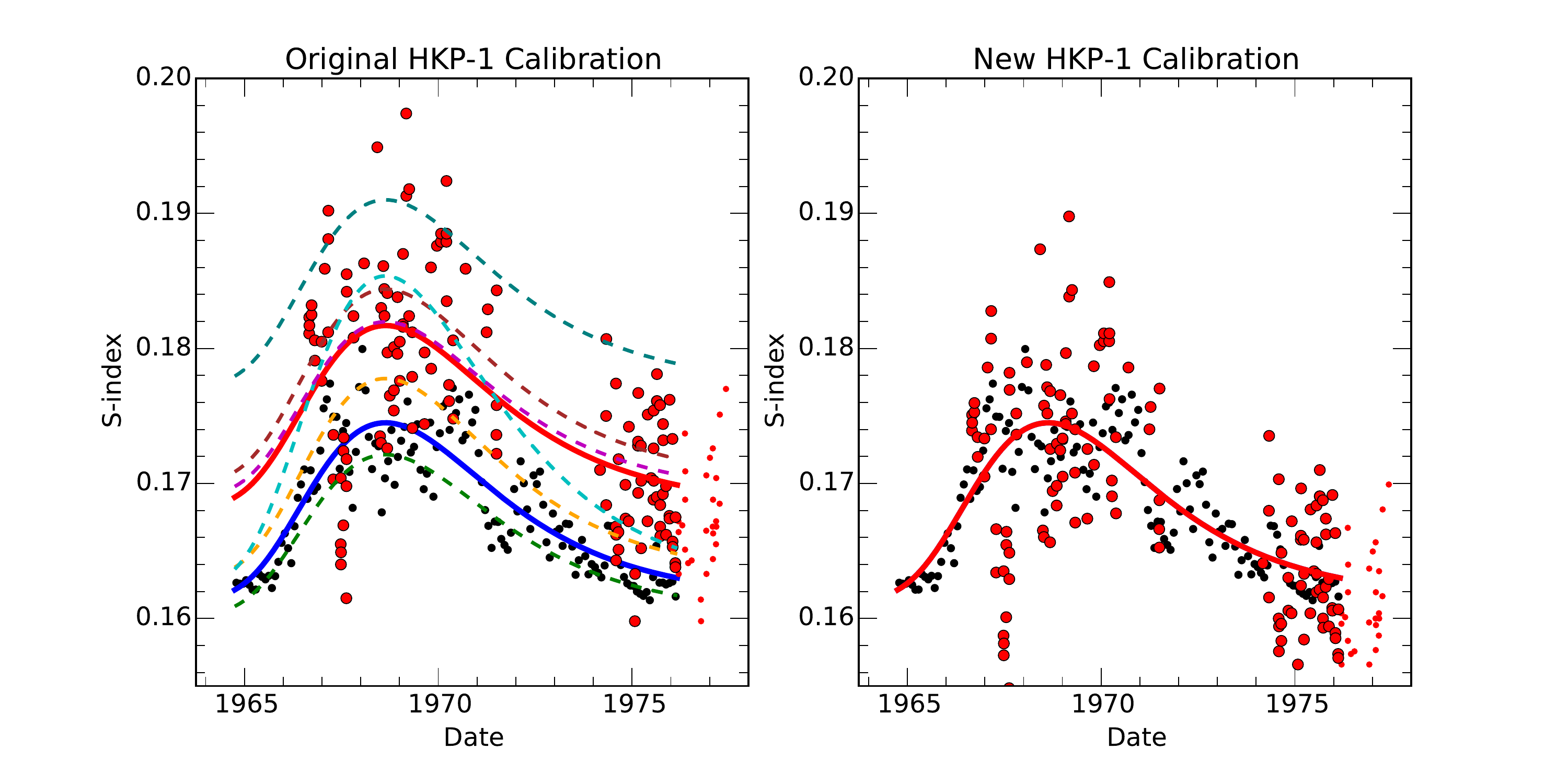}
  \caption{Cycle 20 data and cycle shape model fits.  \emph{Left}: The
    original HKP-1 data are plotted as red circles, with a cycle shape
    model fit to those data as a red line.  Monthly averaged $K_\KKL$
    data transformed to the $S$-index scale using equation
    \eqref{eq:k2s} are shown as black points, and a cycle shape model
    fit to those data as a blue line.  Transformations of the $K_\KKL$
    curve to $S$ using relationships found in the literature (see
    Table \ref{tab:lit_k2s}) are shown as colored curves for
    comparison, using the same color scheme as in Figure
    \ref{fig:cyc23}.  \emph{Right}: HKP-1 data calibrated to the HKP-2
    scale using equation \eqref{eq:1to2}}
  \label{fig:cyc20}
\end{figure*}

\revone{A simple inspection of the HKP-1 data alongside the HKP-2 data
  reveals that the calibration of HKP-1 data to the HKP-2 $S$-index
  scale could be improved.  Notably, the HKP-1 data appear higher than
  the HKP-2 data.  Here we perform a simple analysis analogous to that
  of section \ref{sec:23direct} to illustrate the
  problem.}  Using the cycle boundaries and times of maxima of
  \cite{Hathaway:1999} based on the sunspot record, we
  take the cycle 20 maximum to be 1969/03 and the cycle 20-21 minimum
  to be 1976/03.  We compute the median of a 2-yr window of
the MWO HKP-1 data at these points giving $S_\cmax{20} = 0.182$ (N=35),
and $S_\cmin{21} = 0.168$ (N=54).  Taking the difference we find
$\Delta S_{20} = 0.014$.

The amplitude $\Delta S_{20}$ is slightly less than that of cycle 23, which
agrees with the relative amplitudes of other activity proxies such as
sunspot number and $F_{10.7}$ \citep{Hathaway:2015}.  However, the
directly measured minima is 0.005 $S$-units higher than the cycle 23
value computed in the same way (Section \ref{sec:23direct}) or with
the cycle shape model fit (Section \ref{sec:23fit}).  This
discrepancy, while small in an absolute sense, is over 1/3 of the
cycle amplitude.

The discrepancy in the minima is not unexpected when one considers the
uncertainties involved in calibrating the HKP-1 and HKP-2 instruments.
They were calibrated using near-coincident observations for a sample
of stars resulting in equation \eqref{eq:F2S}, but with individual
stellar ($F$, $S$) means scattered about the calibration curve.
Figure 5 of \cite{Vaughan:1978} shows the calibration data and
regression curve for HKP-1 $F$ and HKP-2 $S$.  Near the solar mean $S$
of $\sim$0.170, scatter about the calibration curve of $\sim$5\% is
apparent.  As a result, for any given star (or the Moon), an
additional correction that in a shift of the mean of about $\sim$5\%
may be required to achieve continuity in the time series.

We apply such a correction to the HKP-1 Moon data to place it on the
same scale as HKP-2.  We use the $K_\KKL$ data as a proxy to tie the
two datasets together, following a similar procedure described in
Section \ref{sec:23fit}.

\revone{
First, we define the boundaries of cycle 20 as 1964/10 to 1976/03 as
in \cite{Hathaway:1999}.  We then fit a cycle shape model curve to the
$K_\KKL$ data in this interval, obtaining the parameters $\{A_\KKL,
t_m, B, \alpha, f_{\ncmin, \KKL} \}$.  Holding the shape parameters
$\{ t_m, B, \alpha \}$ fixed, we fit another curve to the HKP-1 data
to find $\{ A_{1}, f_{\ncmin, 1} \}$.  Transforming the $K_\KKL$ curve
amplitude and offset parameters $\{A_\KKL, f_{\ncmin, \KKL} \}$ to
the HKP-2 scale with equation \eqref{eq:k2s}, we obtain the HKP-2
parameters $\{ A_{2}, f_{\ncmin, 2} \}$.  Finally, using the analog of
equation \eqref{eq:k2s_model_params} for the HKP-1 and HKP-2 amplitude
and offset parameters, we then obtain the transformation:

\begin{equation}\label{eq:1to2}
  S_{\rm HKP-2} = 0.9738 \, S_{\rm HKP-1} - 0.0025
\end{equation}
}

Figure \ref{fig:cyc20} shows the results of this new calibration.  The
left panel shows the HKP-1 data with the original calibration.  Comparing
the red curve (cycle shape model fit to the HKP-1 data) and the blue
curve (fit to the $S(K_\KKL)$ data with equation
\eqref{eq:k2s}), we see clearly the \revone{$\sim$$0.007 \, S$ offset}
with the HKP-2 scale determined above.  Literature relationships from Table
\ref{tab:lit_k2s} are also shown, as applied to a $K_\KKL$ cycle shape
model curve.  Agreement here is generally better than to the HKP-2
data, especially in the case of the \cite{Radick:1998} calibration.
Applying the HKP-1 to HKP-2 transformation of \eqref{eq:1to2} to the
MWO data, the blue and red curves coincide, as shown in the right
panel.

\revone{
The $\sim$$0.007$ offset between the original HKP-1 calibration to $S$
and our HKP-2 calibration demonstrates the principal reason for the
discrepancy between our results and the generally higher values for
$\anglemean{S}$ in previous works summarized in Table
\ref{tab:lit_k2s}.  Without the advantage of HKP-2 measurements of
reflected sunlight from the Moon, previous authors seeking an $S(K)$
relationship using only cycle 20 HKP-1 data for the Sun
\citep{Duncan:1991,Baliunas:1995,Radick:1998} were susceptible to this
systematic offset of $\sim$$0.007 \, S$.  \cite{White:1992}, on the
other hand, used the \cite{Wilson:1978} published data on the $F$
scale.  Coincidentally, those data had a lower value with
$\anglemean{F} = 0.171$ which puts their original $S(K)$ (actually, an
$F(K)$ relationship) estimate closer to ours (see Figure
\ref{fig:cyc23}).  However, the offset error was partially introduced
when they chose to average with the \cite{Duncan:1991} result.  The
remaining differences are due to the myriad of problems associated
with coupling the NSO/KP or SP $K$ measurements to the HKP-1
measurements, either using proxy time series
\citep{White:1992,Baliunas:1995} or stellar observations with the Lick
spectrograph \citep{Duncan:1991,Radick:1998}, which we discussed in
section \ref{sec:prevS}.
}

The scatter of the HKP-1 measurements is somewhat larger than those
from HKP-2.  The reader will also notice a cluster of unusually low
measurements in 1967.  We investigated these points in more detail,
but could not find any anomaly with respect to Moon phase at time of
measurement, or anything in the MWO database that would suggest a
problem with the observations.  With no strong basis for removal of
these observations we keep them in our analysis.

\subsection{Composite MWO and NSO/SP $S(K)$ Time Series}

\begin{figure*}[htb!]
  \centering
  \includegraphics[width=\textwidth]{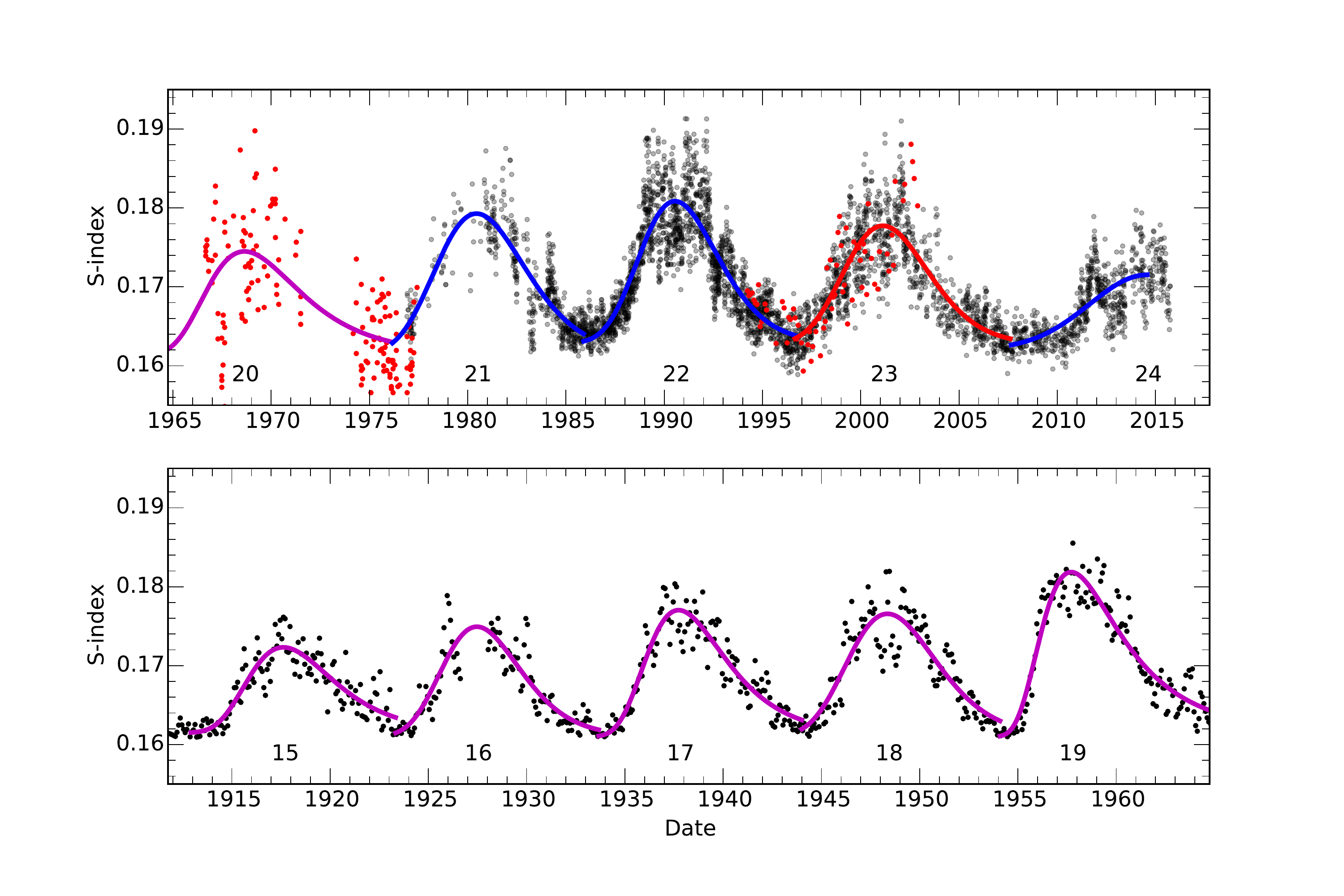}
  \caption{\emph{Top:} composite time series of nightly MWO Moon
    measurements (red) with daily NSO/SP data (black) converted to the
    MWO HKP-2 scale.  \emph{Bottom:} Monthly averaged $K_\KKL$ data
    transformed to the MWO HKP-2 scale.  Cycle shape model curves are
    shown in red when fit using MWO data, blue when fit using only
    NSO/SP data, and magenta when fit using $K_\KKL$ data.  Cycle
    numbers are shown below each cycle.}
  \label{fig:composite}
\end{figure*}

\begin{deluxetable*}{c|cccccc|cccc}
  \tabletypesize{\footnotesize}
  \tablecolumns{11} 
  \tablewidth{0pt}
  \tablecaption{ Cycle Fit Parameters and Measurements \label{tab:cycshape}}
  \tablehead{\colhead{Cycle} & \colhead{$t_{\rm start}$} & \colhead{$t_m$} & \colhead{$B$} & \colhead{$\alpha \times 10^5$} & \colhead{$f_\ncmin$} & \colhead{$A$} & \colhead{$S_\ncmin$} & \colhead{$S_\ncmax$} & $\Delta S_\cyc$ & $\anglemean{S_\cyc}$ }
  \startdata
  15 & 1912.917 & 1917.617 & 2.145 & 7657  & 0.1615 & 0.01083   & 0.1615 & 0.172 & 0.0108 & 0.1671 \\[0.5em]
  16 & 1923.333 & 1927.465 & 1.992 & 5080  & 0.1610 & 0.01391   & 0.1614 & 0.175 & 0.0135 & 0.1678 \\[0.5em]
  17 & 1933.667 & 1937.731 & 1.991 & 9018  & 0.1610 & 0.01604   & 0.1610 & 0.177 & 0.0160 & 0.1690 \\[0.5em]
  18 & 1944.000 & 1948.359 & 2.253 & 4314  & 0.1611 & 0.01552   & 0.1619 & 0.177 & 0.0146 & 0.1696 \\[0.5em]
  19 & 1954.083 & 1957.705 & 1.965 & 12379 & 0.1610 & 0.02086   & 0.1610 & 0.182 & 0.0208 & 0.1716 \\[0.5em]
  20 & 1964.750 & 1968.639 & 2.334 & 7743  & 0.1614 & 0.01314   & 0.1621 & 0.174 & 0.0124 & 0.1684 \\[0.5em]
  21 & 1976.167 & 1980.447 & 2.275 & 3929  & 0.1615 & 0.01778   & 0.1629 & 0.179 & 0.0164 & 0.1717 \\[0.5em]
  22 & 1985.900 & 1990.548 & 2.003 & 4394  & 0.1629 & 0.01801   & 0.1631 & 0.181 & 0.0178 & 0.1713 \\[0.5em]
  23 & 1996.646 & 2001.122 & 2.154 & 3426  & 0.1627 & 0.01504   & 0.1634 & 0.178 & 0.0143 & 0.1701 \\[0.5em]
  24 & 2007.654 & 2014.577 & 2.942 & 1329  & 0.1621 & 0.00945   & 0.1626 & 0.172 & 0.0089 & 0.1670 \\[0.5em]
  \hline\\[-0.5em] 
  $\anglemean{15-24}$ & &  &       &       &        &           & 0.1621 & 0.177 & 0.0145 & 0.1694 \\[0.5em]
  $\sigma_{\rm measure}$ & & &     &       &        &           & \revone{0.0008} & 0.001 & \revone{0.0012} & 0.0005 \\[0.5em]
  $\sigma_{\rm scatter}$ & & &     &       &        &           & 0.0008 & 0.003 &  0.003 & 0.002
  \enddata
\end{deluxetable*}

We have now calibrated both the NSO/SP data and the MWO HKP-1 data to
the HKP-2 scale.  The complete composite time series covering cycles
20--24 is shown in top panel of Figure \ref{fig:composite}.  The KKL
composite \citep{Bertello:2016} allows us to further
extend $S$ back to cycle 15, as shown in the bottom panel the
figure.  Note that the KKL data are monthly means, while the NSO/SP
and MWO series are daily measurements.  For each cycle we have
determined the cycle duration using the absolute minimum points of a
1-yr median filter on the NSO/SP data (cycles 21--24) or using
\cite{Hathaway:1999} values (cycles 15--20).  We have fit each cycle
with a cycle shape model (equation \eqref{eq:cycshape}), with the best
fit parameters shown in the left portion of Table \ref{tab:cycshape}.
Cycle 24 is a special case.  Because we only have observations for
half of the cycle, the optimizer has difficulty obtaining a reasonable
fit.  This problem was resolved by constraining $\alpha$ to be a
function of $t_m$ using the relationship found in \cite{Du:2011}.
From these curve fits, we determine the cycle minimum, maximum, max -
min amplitude, and mean value.  These results are summarized in the
right portion of Table \ref{tab:cycshape} and represent our best
estimate of chromospheric variability through the MWO $S$-index over
ten solar cycles.

The uncertainty in the cycle minima and maxima for cycle 23 were found
to be $\approx$ 0.5\% for cycle 23 (section \ref{sec:23fit}), which
are summed in quadrature along with the covariance term give the
uncertainty in the amplitude of $\approx$ 10\%.  These uncertainties
are then propagated into equation \eqref{eq:k2s} which determines the
uncertainties of the other cycles.  The uncertainty in the cycle means
is about 0.3\%, as determined by the Monte Carlo experiment (see
Appendix A).  These relative uncertainties are applied to the cycle
15--24 mean in Table \ref{tab:cycshape} to compute $\sigma_{\rm
  measure}$, the typical uncertainty cycle measurements on the
$S$-index scale.  The standard deviation of the minima, maxima,
amplitudes and means are given as $\sigma_{\rm scatter}$.

\subsection{Conversion to $\log(\RpHK)$}

The $S$-index reference bands $R$ and $V$ (equation \eqref{eq:S}) vary
with stellar surface temperature (and metalicity; see \citep{Soon:1993b}), and furthermore
temperature-dependent flux from the photosphere is present in the $H$
and $K$ bands.  These effects lead to a temperature dependence or
``color term'' in $S$ which limits its usefulness when comparing stars
of varied spectral types.  The activity index $\RpHK$
(\cite{Noyes:1984}) seeks to remove the aforementioned color dependence
of $S$ and is widely used in the literature.  We therefore compute
$\log(\RpHK)$ of the Sun for the purpose of inter-comparison with
stellar magnetic activity variations of other Sun-like stars from the
MWO HK project or elsewhere.  In Table \ref{tab:RpHK} we used the procedure
of \cite{Noyes:1984} to calculate $\log(\RpHK)$ from the $S$
measurements of Table \ref{tab:cycshape}.  In this calculation, we
adopted the solar color index $(B - V)_\Sun = 0.653 \pm 0.003$
\citep{Ramirez:2012}.  The typical uncertainty of the
cycle measurements on the $\log(\RpHK)$ scale, $\sigma_{\rm measure}$, and
standard deviation for the ten cycles, $\sigma_{\rm scatter}$, are
also presented in Table \ref{tab:RpHK}.  The cycle amplitude
expressed as $\log(\Delta\RpHK)$, and the fractional amplitude $\Delta
\RpHK/\anglemean{\RpHK}$ have been used in stellar amplitude studies
in the literature \citep{Soon:1994,Baliunas:1996b,Saar:2002}.

\begin{deluxetable*}{c|ccccc}
  \tabletypesize{\footnotesize}
  \tablecolumns{6} 
  \tablewidth{0pt}
  \tablecaption{ Cycle Measurements in $\log(\RpHK)$ \label{tab:RpHK}}
  \tablehead{\colhead{Cycle} & \colhead{$\log(R'_{\rm HK,min})$} & \colhead{$\log(R'_{\rm HK,max})$} & \colhead{$\log(\Delta\RpHK)$} & \colhead{$\log(\anglemean{\RpHK})$} & \colhead{$\Delta \RpHK/\anglemean{\RpHK}$}}
  \startdata
  15                     & -4.9882 & -4.927 & -5.807 & -4.9552 & 0.141 \\[0.5em]
  16                     & -4.9885 & -4.913 & -5.712 & -4.9513 & 0.174 \\[0.5em]
  17                     & -4.9909 & -4.903 & -5.638 & -4.9443 & 0.203 \\[0.5em]
  18                     & -4.9856 & -4.905 & -5.676 & -4.9413 & 0.184 \\[0.5em]
  19                     & -4.9909 & -4.879 & -5.523 & -4.9305 & 0.256 \\[0.5em]
  20                     & -4.9846 & -4.916 & -5.748 & -4.9481 & 0.158 \\[0.5em]
  21                     & -4.9797 & -4.891 & -5.626 & -4.9300 & 0.201 \\[0.5em]
  22                     & -4.9783 & -4.884 & -5.592 & -4.9319 & 0.219 \\[0.5em]
  23                     & -4.9763 & -4.899 & -5.686 & -4.9385 & 0.179 \\[0.5em]
  24                     & -4.9811 & -4.931 & -5.892 & -4.9558 & 0.116 \\[0.5em]
  \hline\\[-0.5em] 
  $\anglemean{15-24}$    & -4.9844 & -4.905 & -5.690 & -4.9427 & 0.183 \\[0.5em]
  $\sigma_{\rm measure}$ &  \revone{0.0087} &  0.008 & \revone{0.068} &  0.0072 & 0.032  \\[0.5em]
  $\sigma_{\rm scatter}$ &  0.0049 &  0.016 & 0.101 &  0.0093 & 0.038
  \enddata
\end{deluxetable*}

\section{Linearity of $S$ with $K$}
\label{sec:ks_linearity}

In the above analysis we assumed that the ratio of 1 \AA{} emission in
the cores of the H \& K lines, $H$ and $K$ are linearly related such
that $S$ (equation \eqref{eq:S}) and $K$ are also linear as in
equation \eqref{eq:k2s}.  It is also assumed that the pseudo-continuum
bands $V$ and $R$ are constant.  These assumptions were also implicit
in the derivation of $S(K)$ relationships in the literature shown in
Table \ref{tab:lit_k2s}, but we are not aware of any observational
evidence in support of them.

We use emission indices from the SOLIS/ISS instrument ($K$, $H$) and
SSS ($K$, $H$, $V$, $R$) to examine the trends with respect to the $K$
band.  SOLIS/ISS emission indices are derived from spectra normalized
to a single reference line profile obtained with the NSO Fourier
Transform Spectrometer as described in \cite{Pevtsov:2014}.  SSS
indices are derived from continuum normalized and wavelength
calibrated spectra with intensity points at 3909.3770 \AA{} set to
0.83 and 4003.2688 \AA{} at 0.96, and the rest of the spectrum
normalized according to the line defined by those points.

\begin{figure}[htb!]
  \centering
  \includegraphics[width=\columnwidth]{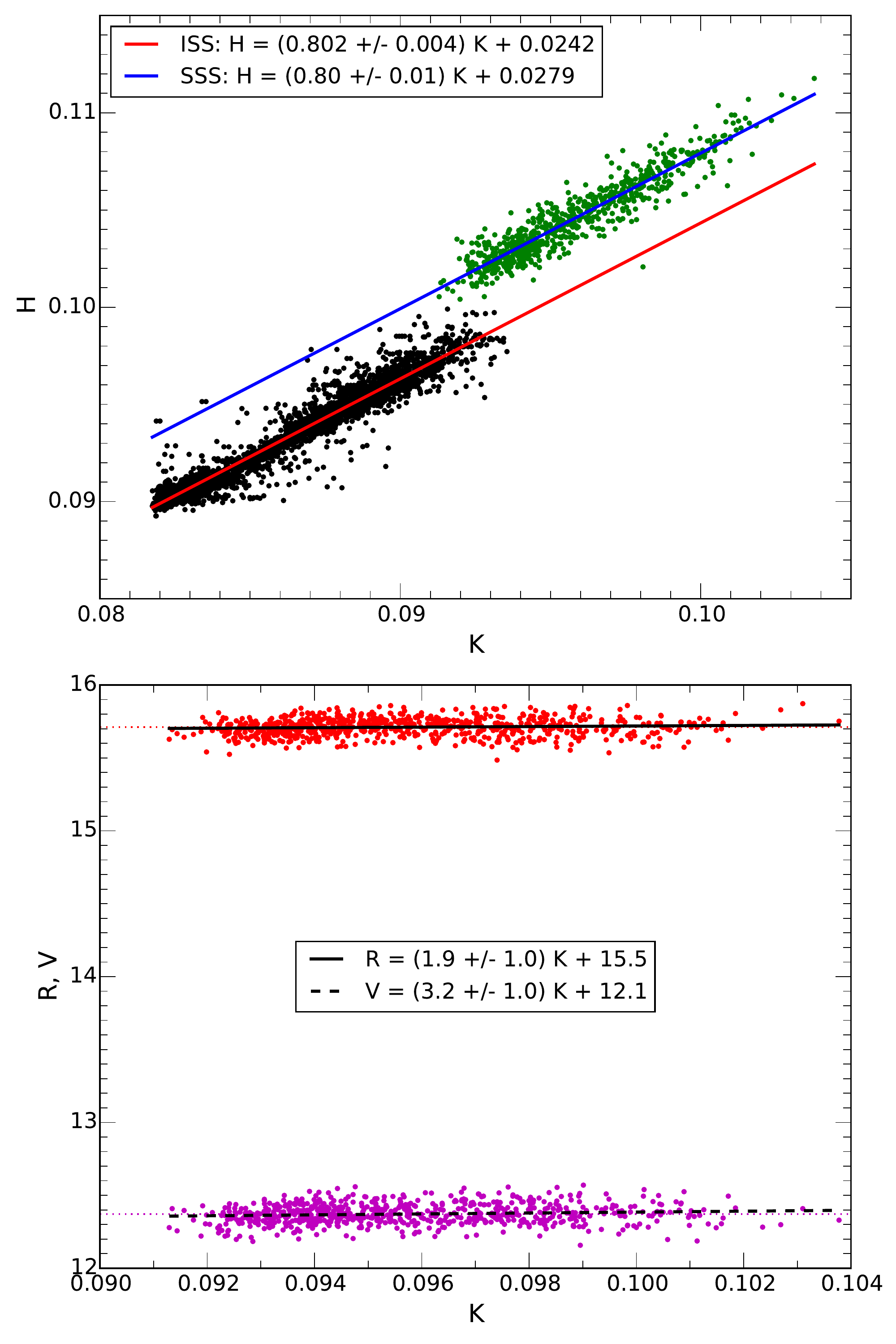}
  \caption{\emph{Top}: Relationship between calcium H \& K 1-\AA{}
    emission indices in SOLIS/ISS (black) and SSS (green).
    \emph{Bottom}: $V$ and $R$ band emission indices versus $K$ from
    SSS spectra.}
  \label{fig:HVRvK}
\end{figure}

Figure \ref{fig:HVRvK} (top) shows the relationship between the
1-\AA{} emission index in the H and K line cores from the NSO
SOLIS/ISS and Lowell SSS instruments.  SOLIS/ISS data (black points)
are from the beginning of observations until the middle of 2015, when
the instrument moved from Kitt Peak to Tuscon and resulted in a
discontinuous shift in the $K/H$ time series which is still under
investigation.  SSS data (green points) includes only the observations
after the camera was upgraded in 2008, which is significantly less
noisy than with the previous CCD.  Differences in resolution and
systematics such as stray light are likely responsible for the offsets
in both $K$ and $H$ between the two instruments.  However, in both
instruments the $H(K)$ relationship is linear with a slope $b < 1$ and
a zero point $a > 0$.  The slopes found from each instrument,
$\approx$ 0.8, agree to within the uncertainties.  Linearity of $H$
with $K$ is compatible with the assumption of linearity of $S$ with
$K$.  The $H(K)$ linearity further implies that the emission ratio
$K/H$ has the form:

\begin{equation}
  \frac{K}{H} = \frac{K}{a + bK}
\end{equation}

With nonzero $a$ the ratio must vary over the solar cycle.  This ratio
is a diagnostic of the optical depth of the surface integrated
chromosphere \citep{Linsky:1970}.  The data show that the ratio
increases by $\sim 2$\% over the rising phase of solar cycle 24.

Figure \ref{fig:HVRvK} (bottom) shows $V$ (3891.0--3911.0 \AA{}) and
$R$ (3991.0--4011.0) \AA{} integrated emission indices versus $K$
from continuum normalized SSS solar spectra.  The data show these 20
\AA{} pseudo-continuum bands to be nearly constant with activity, with
the $V$ and $R$ bands having RMS variability of 0.6\% and 0.4\%,
respectively.  This is in agreement with the small variance in the
$C_{\rm RV}$ index based on $V$ and $R$ found by \cite{Soon:1993b}.
Linear regression of the $V(K)$ and $R(K)$ gives uncertainties in the
slope (see Figure \ref{fig:HVRvK}) that place them $3.2\sigma$ and
$1.9\sigma$ from zero, respectively, giving them marginal statistical
significance.  The relative increase in flux indicated by these slopes
for $V(K)$ and $R(K)$ over the full range of $K$ are 0.3\% and 0.1\%
respectively.

To conclude, the data show $H$ to be linear with $K$, and $V$ and $R$
to be nearly constant, which is compatible with the model that $S = a
+ b K$ as assumed in the previous sections.

\section{Conclusion}
\label{sec:conclusion}

We have used the observations of the Moon with the MWO HKP-2
instrument to accurately place solar cycle 23 on the $S$-index scale.
By deriving a proxy with the NSO/SP $K$-index data we extend the solar
record to include cycles 21--24.  We found that our cross-calibration
method using a cycle shape model has lower uncertainty than averaging
and regression methods when the number of observations is small.  The
Kodaikanal Observatory plage index data calibrated to the Ca K 1-\AA{}
emission index were used to calibrate the MWO HKP-1 measurements of
cycle 20 to the HKP-2 scale, as well as to extend the $S$-index record
back to cycle 15.  The full composite time series from the KKL, HKP-1,
NSO/SP, and HKP-2 instruments forms a record of chromospheric
variability of the Sun over 100 years in length which may be compared
to the stellar observations of the MWO and SSS programs, as well as
other instrumental surveys which have accurately calibrated their data
to the HKP-2 $S$ scale.

We find a mean value of $\anglemean{S} = 0.1694 \pm 0.0005$ for the
Sun that is 4\%--9\% lower than previous estimates in the literature
(see Table \ref{tab:lit_k2s}) that used MWO HKP-1 data or stellar
observations for their calibration.  \revone{We believe the
  discrepancy is due to (1) a systematic $+0.007$ offset in
  the previous $S$ calibration of the HKP-1 solar data that is within
  the scatter of the $F$ and $S$ relationship of \cite{Vaughan:1978},
  and (2) uncertainties introduced in coupling those HKP-1 measurements to the
  non-overlapping NSO/KP and SP $K$ time series using either proxy
  activity time series or stellar measurements with the Lick
  spectrograph.}  This relatively small change in
$S$ on the stellar activity scale (where $S$ ranges from 0.13 to 1.4
\citep{Baliunas:1995}) is a rather large fraction of the cycle
amplitude, which we estimate to be $0.0145 \pm 0.0014$ in $S$, on
average.  \revone{Our results are consistent with previous estimations
  from the SSS \citep{Hall:2004,Hall:2007b,Hall:2009}, as well as our
  new reduction of that dataset (see section \ref{sec:sss}).  Our
  results are also consistent with parallel work by \cite{Frietas:2016},
  who calibrate HARPS Ca \II{} H \& K observations to the MWO
  $S$-index scale using an ensemble of solar twins, and found
  $\anglemean{S} = 0.1686$ for the Sun using observations of asteroids
  during cycles 23 and 24.}

We do not expect our change in solar $S$ to have large consequences on
previous studies that used the Sun \revone{as just another star} in
stellar ensembles covering a wide range of rotation periods and
spectral types.  \revone{However, studies which used the Sun as an
  absolutely known anchor point in activity relationships could be
  significantly affected by this shift \citep[e.g.][who compromises
    and adopts the mean of Baliunas' and Hall's solar $S$-index in
    their activity-age relationship]{Mamajek:2008}.}  Detailed
comparisons of the Sun to solar twins will \revone{also} be sensitive
to this change.  For example, MWO HKP-2 observations of famous solar
twin 18 Sco from 1993--2003 gives a mean $S$-index of 0.173, which we
find to be $\sim$2\% higher than the Sun, but using other estimates
from the literature we would conclude that the activity is 2\%--3\%
lower than the Sun.  This star has a measured mean rotation period of
22.7 days \citep{Petit:2008}, slightly faster than the Sun, which
would lead us to expect a higher activity level if in all other
respects the star really is a solar twin.

Cycle amplitude is another quantity that can be compared to stellar
measurements to put the solar cycle in context.  So far as Ca \II{} H
\& K emission is proportional to surface magnetic flux, this gives an
estimate of how much the surface flux changes over a cycle period.
This $d B / d t$ is the surface manifestation of the induction
equation at the heart of the dynamo problem.  \cite{Soon:1994} was the
first to study this for the Sun and an ensemble of stars using the
fractional amplitude $\Delta \RpHK / \anglemean{\RpHK}$.  They found
an inverse relationship between fractional amplitude and cycle period,
which is also seen in the sunspot record.  The fractional solar
amplitude measured here has a mean value of $0.18 \pm 0.03$ and
ranges from 0.116 (cycle 24) to 0.255 (cycle 19).  Our fractional
amplitudes for cycles 21 and 22 are close to the value of 0.22 found
in \cite{Soon:1994} using NSO/KP data and the \cite{White:1992} $S(K)$
transformation, and our range over cycles 15--24 largely overlaps with
their range of values 0.06--0.17 found by transforming the sunspot
record to $S$.  \cite{Egeland:2015} recently measured four cycle
amplitudes for the active solar analog ($(B-V) = 0.632$) HD 30495,
with fractional amplitudes $\Delta \RpHK / \anglemean{\RpHK}$ ranging
from 0.098 to 0.226 comparable to our solar measurements despite the
2.3 times faster rotation of this star.  However, when using absolute
amplitudes the largest HD 30495 cycle has $\Delta S = 0.047$, which is
2.3 times the \emph{largest} solar amplitude of 0.0211 for cycle 19.
This illustrates that the use of fractional amplitudes obscures the
fact that the more active, faster rotators in general have a much
larger variability than the Sun, which indicates a much more efficient
dynamo.  Indeed, \cite{Saar:2002} studied cycle amplitudes for a
stellar ensemble and found that $\Delta \RpHK / \anglemean{\RpHK}
\propto \anglemean{\RpHK}^{-0.23}$, fractional amplitude
\emph{decreases} with increased activity.  In an upcoming work we will
use the longer timeseries available today to reexamine cycle
properties such as amplitude and period for an ensemble of solar
analogs.

Our value of $S$ is significantly higher than the basal flux estimate
of \cite{Livingston:2007} of $0.133 \pm 0.006$ using the center disk H
+ K index values from NSO/KP, transformed to the $S$-index scale using
the flux relationships of \cite{Hall:2004}.  The center disk
measurements integrate flux from a small, quiet region near disk
center where little or no plage occurs, and the derived $S$ estimate
purported to be indicative of ``especially quiescent stars, or even the
Sun during prolonged episodes of relatively reduced activity, as
appears to have occurred during the Maunder Minimum period''.  If the
latter assumption were true, we estimate it would reflect an 18\%
reduction in $S$ from current solar minima to Maunder Minimum, or
about twice the amplitude of the solar cycle.  Total solar irradiance
varies by $\sim$0.1\% over the solar cycle \citep{Yeo:2014}, so
further assuming a linear relationship between $S$ and total solar
irradiance this would translate into a 0.2\% reduction in flux at the
top of the Earth's atmosphere.  The validity of these assumptions are
uncertain.  Precise photometric observations of a star transitioning
to or from a flat activity phase would greatly aid in determining the
relationship between grand minima in magnetic activity and irradiance.

\cite{Schroder:2012} measured the $S$-index for the Sun from daytime
sky observations using the Hamburg Robotic Telescope (HRT) during the
extended solar minimum of cycle 23--24 (2008--2009).  Their
instrumental $S$-index was calibrated to the MWO scale using 29 common
stellar targets and published MWO measurements.  They report an
average $\anglemean{S} = 0.153$ over 79 measurements during the
minimum period, and discuss their absolute minimum $S$-index of 0.150
on several plage-free days, comparing this ``basal'' value to the
activity of several flat-activity stars presumed to be in a Maunder
Minimum-like state.  Our NSO/SP $S(K)$ proxy covers the same minimum
period and has an absolute minimum measurement of 0.1596 on 9 Oct
2009, while the lowest Moon measurement from the HKP-2 instrument is
0.1593 on 28 Jan 1997.  Inspection of SOHO MDI magnetograms on those
dates shows no significant magnetic features on the former date, while
one small active region is present in the latter.  The uncertainty in
the calibration scale parameter from $S_{\rm HRT}$ to $S_{\rm MWO}$
(HKP-2) is about 2\% \cite{Schroder:2012}, making their absolute
minimum $S$ $\sim 2 \sigma$ below ours.  \cite{Baliunas:1995} found a
systematic error in the value and amplitude of the solar $S$-index
measured from daytime sky observations, which they attributed to
Rayleigh scattering in the atmosphere, ultimately deciding to omit
those observations from their analysis.  While \cite{Schroder:2012}
applied a correction for atmospheric scattering and estimated a small
0.2--1.8\% error for it, we suspect that some systematic offset due to
scattering remains that explains their lower $S$-index measurements.

Our results compare favorably with the independently calibrated SSS
instrument at Lowell Observatory.  While the MWO, NSO/SP, and NSO/KP
programs have ceased solar observations, SSS continues to take
observations of Ca H \& K, as does the SOLIS/ISS instrument which
began H \& K-line observations in Dec 2006 \citep{Bertello:2011}.  Combining
equations \eqref{eq:sac2iss} and \eqref{eq:k2s}, we obtain $S(K_{\rm
  ISS}) = 1.71 \, K_{\rm ISS} + 0.02$, which can be used to transform
data on the SOLIS/ISS scale to $S$, including the composite $K$-index
dataset from 1907--present \citep{Bertello:2016,Data:Kcomposite}.

This work illustrates once again the complexities of comparing and
calibrating data on several different instrumental flux scales.  Some
of this confusion could be avoided if more effort were  put toward
calibrating instruments to physical flux (\fluxunits{}).  Should this be
done, discussions of discrepancies would be more about the validity of
the methods used to achieve the absolute calibration rather than the
details of the chain of calibrations used to place measurements on a
common scale.  We believe the former path is preferable, though not
without its own substantial difficulties.  Given the encouraging
agreement between the SSS and MWO HKP-2 solar and stellar data, the
$S$ to flux relationships presented in \cite{Hall:2007b} are a good
starting point for placing the MWO data on an absolute scale.  Further
work in this area should carefully evaluate the $S$ to flux
relationship, so that calibrated Ca \II{} H \& K flux measurements by
future instruments may immediately be comparable to the extensive and
pioneering Mount Wilson Observatory observations.

\vspace{1em}
Thanks to Phil Judge, Giuliana de Toma, Doug Duncan, Robert Donahue,
and Alfred de Wijn for the useful discussions which contributed to
this work.  Thanks to Steven Keil and Tim Henry for calibrating and
providing NSO data products.  R.E. is supported by the Newkirk
Fellowship at the NCAR High Altitude Observatory.  W.S. is supported
by SAO grant proposal ID 000000000003010-V101.  J.C.H. thanks Len
Bright, Wes Lockwood and Brian Skiff for obtaining most of the SSS
solar data over the years, and Len Bright for his ongoing curation of
the SSS solar and stellar data.  A.A.P. acknowledges the financial
support by the Academy of Finland to the ReSoLVE Centre of Excellence
(project no. 272157).

\revone{
\section*{Appendix A: Monte Carlo Cycle Model Fitting Experiments}

\begin{figure*}[htb!]
  \centering
  \includegraphics[width=\textwidth]{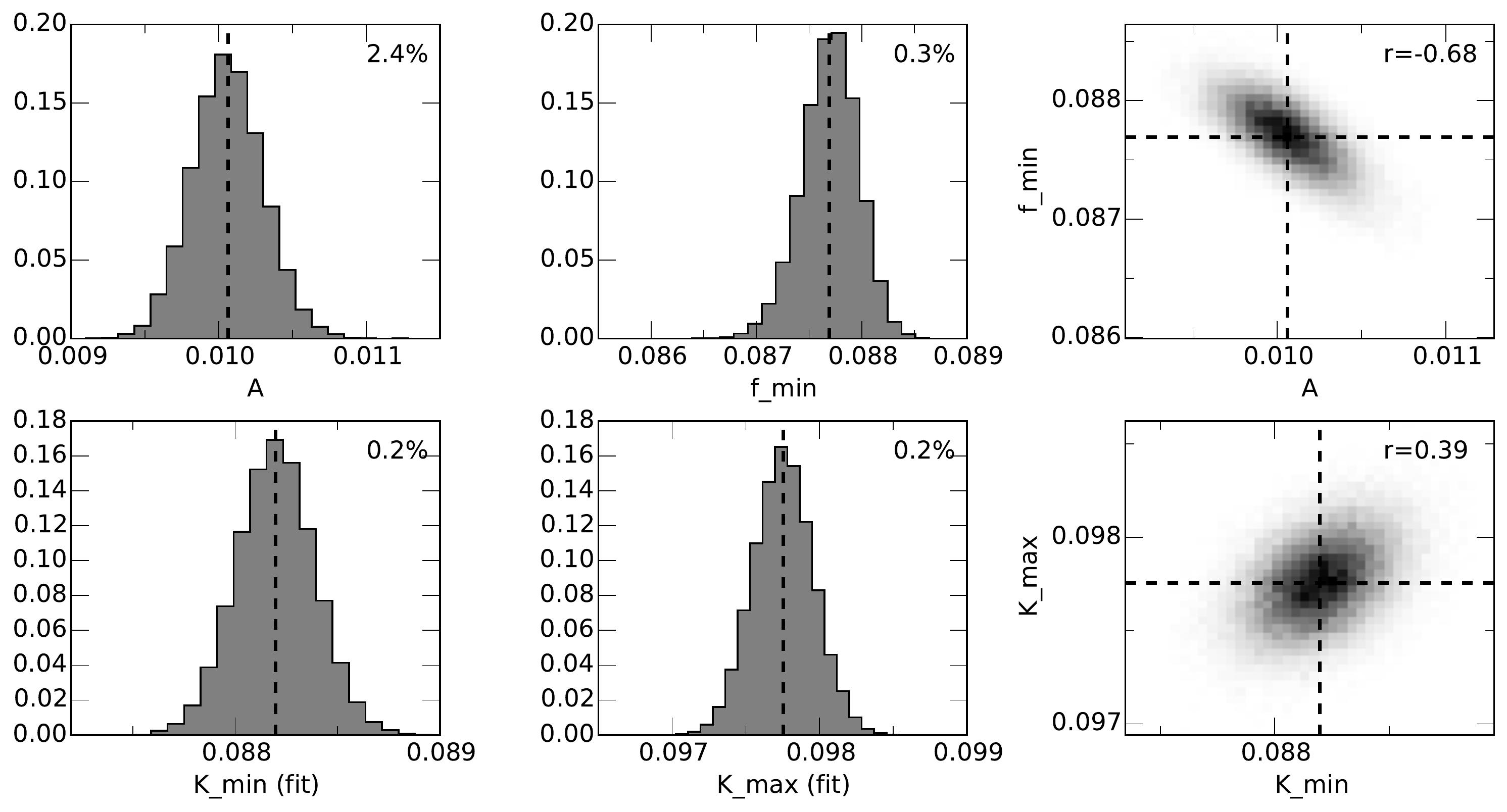}
  \caption{Monte Carlo experiment to determine the uncertainty in
    cycle model amplitude and offset parameters, $\{A, f_\ncmin\}$
    (top row) and the cycle minimum and maximum $\{K_\ncmin,
    K_\ncmax\}$ (bottom row) using the NSO/SP $K$-index time series
    for cycle 23.  Distributions show results of Monte Carlo trials
    sampling 80\% of the 1087 measurements in cycle 23.  The standard
    deviation of each 1D distribution is shown as a percentage in the
    top right corner.  The correlation coefficient $r =
    \sigma_{xy}/\sigma_x \sigma_y$ is shown in the top right corner of
    the 2D distributions of the right column.  Each histogram is
    normalized by the number of trials.  The ``true'' value from a fit
    using all measurements is shown with a dashed line.}
  \label{fig:mc1}
\end{figure*}

\begin{figure*}[htb!]
  \centering
  \includegraphics[width=\textwidth]{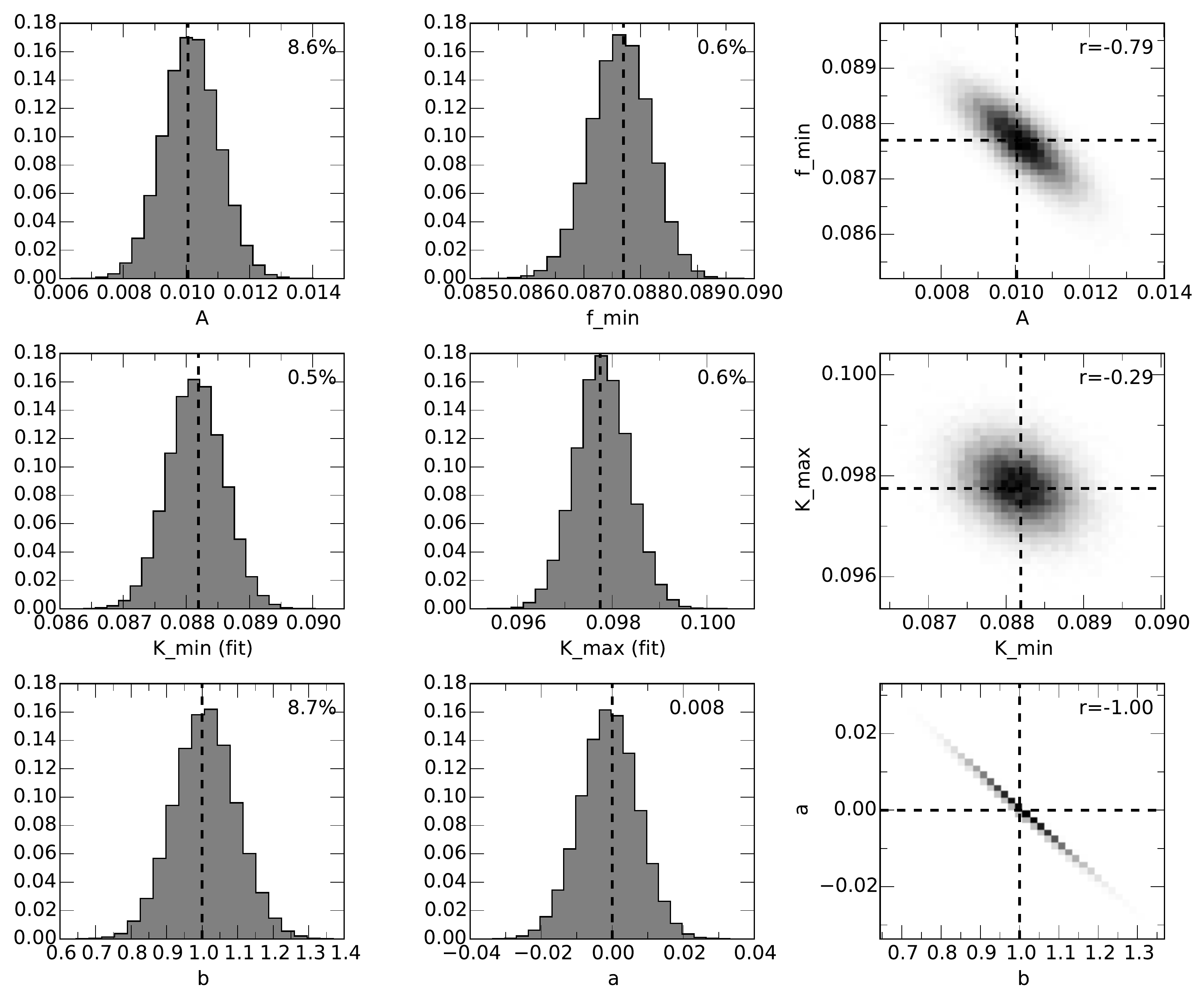}
  \caption{Monte Carlo experiment to determine the uncertainty in the
    scale factor of $S(K)$ using only 56 points randomly drawn from the
    NSO/SP data for cycle 23.  \emph{Top row:} cycle shape model
    parameters $A$ and $f_\ncmin$.  \emph{Middle row:} cycle minima
    and maxima determined by the model fit.  \emph{Bottom row:} linear
    transformation parameters for $K_{\rm true} = a + b K_i$ from the trial
    measurements $K_i$ to the true scale, $K_{\rm true}$.  Each histogram is
    normalized by the number of trials.  The ``true'' value from a fit
    using all measurements is shown with a dashed line.}
  \label{fig:mc2}
\end{figure*}

In section \ref{sec:23fit} we used fits to the NSO/SP
data and 56 MWO HKP-2 data points during the rising phase of cycle 23 to
determine the minimum, maximum, amplitude and average value on the
HKP-2 $S$-index scale.  These fits were again used in section
\ref{sec:k2s} to determine the $S(K)$ relationship of equation
\eqref{eq:k2s}.  We determine the uncertainty in these measurements
using two Monte Carlo experiments described here.

The first experiment is aimed at understanding the uncertainties in
the cycle shape fit parameters when fitting the NSO/SP $K$-index data.
First, we determine the limits of cycle 23 using a 1-yr median filter
on the data and taking the absolute minimum points before and after
the maximum.  There are 1087 NSO/SP measurements in this period.  In
each trial we select with replacement 80\% of the measurements (N=869)
and fit the cycle shape model (equation \ref{eq:k2s}) parameters $\{A,
t_m, B, \alpha, f_\ncmin\}$ to those data using the TRR+LM algorithm.
In addition to the fit parameters we measure the rise phase minimum
(beginning of the cycle model) and maximum value, $\{K_\ncmin,
K_\ncmax\}$.

We ran 50,000 Monte Carlo trials.  The distributions for the amplitude
and offset parameters, $\{A, f_\ncmin\}$, and the cycle minima and
maxima are shown in Figure \ref{fig:mc1}.  We computed the correlation
coefficients ${\rm cov}(x, y) / (\sigma_x \cdot \sigma_y)$ for all five
model parameters plus the two cycle measurements, obtaining the
following symmetric correlation matrix:

\[
\begin{blockarray}{rrrrrrr}
   A     &  t_m   & B      & \alpha & f_\ncmin & K_\ncmin & K_\ncmax \\
\begin{block}{[rrrrrrr]}
   1     &  0.493 &  0.303 & -0.158 & -0.684 & -0.690 &  0.280 \\
         &  1     &  0.314 & -0.122 & -0.933 & -0.753 & -0.645 \\
         &        &  1     & -0.846 & -0.450 & -0.710 & -0.234 \\
         &        &        &  1     &  0.265 &  0.463 &  0.163 \\
         &        &        &        &  1     &  0.914 &  0.510 \\
         &        &        &        &        &  1     &  0.389 \\
         &        &        &        &        &        &  1     \\
\end{block}
\end{blockarray}
\]

The amplitude $A$ has a high negative correlation with the minima
parameters, $f_\ncmin$ and $K_\ncmin$, indicating that high minima are
compensated with low amplitudes, such that the maxima point is not too
large.  Indeed, there is a low correlation between $A$ and $K_\ncmax$.
In general, the minima measurement $K_\ncmin$ is more highly
correlated with the fit parameters than the maxima measurement
$K_\ncmax$ The high correlation between the time of maximum $t_m$ and
$f_\ncmin$ indicates that fits with early maxima have a lower minima.
We are not interested in the correlations among the shape parameters
${t_m, B, \alpha}$ as they involve time scales in the cycle which are
not studied in this work.

The percent standard deviation of each quantity $x$ is
$\mathbf{\sigma_x/\anglemean{x}}/100 = (2.4, 4.2, 19, 6.7, 0.31, 0.22,
0.21)$, with the same ordering as in the above matrix.  For $t_m$, we
calculated $\sigma$ relative to 1 year instead of the full decimal
year of maximum.  We find that the standard deviation is quite small
for the quantities of most concern: $\{A, f_\ncmin, K_\ncmin,
K_\ncmax\}$.

In our second Monte Carlo experiment we determine the uncertainty in
fitting a partial cycle using relatively few data points, as was done
for the MWO HKP-2 data for cycle 23.  We take the fit to the full
NSO/SP $K$-index dataset (N=1087) in cycle 23 to be the ``true''
cycle defined by the parameters $\{A, t_m, B, \alpha, f_\ncmin \}$.
We then ran 50,000 Monte Carlo trials in which we randomly selected
only 56 data points from the rise phase of the cycle, up to the time
of the last MWO HKP-2 measurement.  The selections are drawn from
bins according to an N=10 equal-density binning of the MWO data points
in order to ensure each trial maintained a sampling relatively uniform
in time.  In each trial, we randomly draw a set of cycle shape
parameters $\{t_m, B, \alpha\}$ from the previous Monte Carlo
experiment.  This is done in order to incorporate the uncertainty of
the shape parameters into our results.

We use the same TRR+LM fitting procedure to find the remaining
parameters $\{ A, f_\ncmin \}$.  We compute the cycle minimum and
maximum $\{K_\ncmin, K_\ncmax \}$ from each model fit.  As a check on
our uncertainty derivation in equation \eqref{eq:k2s_model_error} we
also compute distributions of linear fit parameters $\{a_i, b_i\}$
using equation \eqref{eq:k2s_model} where the ``true'' values are
$\{0, 1\}$.

The results from the second experiment are shown in Figure
\ref{fig:mc2}.  The percent standard deviation for the parameters
$\{A, f_\ncmin, K_\ncmin, K_\ncmax\}$ are $\{8.6, 0.60, 0.50, 0.58\}$.
We find that using only 56 data points during the rise phase increases
the uncertainty in $A$ by a factor of 3.6 compared to the previous
experiment.  The offset parameter $f_\ncmin$ is more robust, with the
uncertainty increasing only by $\sim 65$\%.  The relative standard
deviation for the amplitude $\Delta K = K_\ncmax - K_\ncmin$ was 8.4\%, and that
of the cycle mean $\anglemean{K}$ was 0.29\%.  In section \ref{sec:23fit}, we use the
relative standard deviations found in this experiment to estimate the
uncertainty of $S_\ncmin$, $S_\ncmax$, $\Delta S$, and $\anglemean{S}$
from the 56 HKP-2 measurements of cycle 23.
}

\section*{Appendix B: Observations}

\begin{deluxetable}{ccccccc}[ht!]
\tabletypesize{\small}
\tablewidth{0pt}
\tablecaption{Calibrated Solar $S$-index Time Series \label{tab:data}}
\tablehead{
\colhead{Date} & \colhead{Original} & \colhead{Calibrated} &\colhead{Instrument} \\
\colhead{(JD $-$ 2,400,000)} & \colhead{$S$ or $K$} & \colhead{$S$} & \colhead{}
}
\startdata
39370.857 &  0.177315 & 0.173915 & MWO/HKP-1 \\
43103.208 &  0.088529 & 0.163954 & NSO/SP    \\
49438.828 &  0.169500 & 0.169500 & MWO/HKP-2 \\
49454.496 &  0.164500 & 0.159725 & SSS/CCD-1 \\
54679.073 &  0.163200 & 0.163200 & SSS/CCD-2 \\
\vdots    &  \vdots   & \vdots   & \vdots
\enddata
\tablecomments{Table \ref{tab:data} is presented in its entirety in the electronic 
edition of the Astrophysical Journal.  A portion is shown here for
guidance 
regarding its form and content.}
\end{deluxetable}

We provide the daily observations of solar Ca \II{} K or HK emissions
used in this work.  An example of the data are shown in Table
\ref{tab:data}.  Data from five instruments are provided, denoted
MWO/HKP-1, MWO/HKP-2, NSO/SP, SSS/CCD-1, SSS/CCD-2. We provide both
the original calibration of the data, as an $S$-index or as the
$K$-index in the case of NSO/SP, as well as the calibration to the
MWO/HKP-2 scale described in this work.  These data may be used to
recreate Figures \ref{fig:cyc23}, \ref{fig:cyc20}, \ref{fig:composite}
(top panel), and \ref{fig:sss}.  The KKL--NSO/SP--ISS composite
\citep{Bertello:2016} used in the bottom panel of Figure
\ref{fig:composite} are not included here, but they are publicly
available from the Harvard Dataverse \citep{Data:Kcomposite}.
Observation times are given as a modified Julian Date, and are on the
UTC time scale.

\pagebreak

\bibliography{mwosun}
\bibliographystyle{apj}

\end{document}